\documentstyle[epsfig]{mn2e}
\begin{document}
\input ./BoxedEPS.tex
\SetEPSFDirectory{./}
\SetRokickiEPSFSpecial
\HideDisplacementBoxes

\title
[
Sub-mm observations of the Hubble Deep Field and Flanking Fields]
{Sub-millimetre observations of the Hubble Deep Field and Flanking Fields}

\author[S. Serjeant, J.S. Dunlop, R. Mann, et al.]{
S. Serjeant$^{1,2}$, 
J.S. Dunlop$^3$, 
R. G. Mann$^{3}$, 
M. Rowan-Robinson$^{1}$, 
D. Hughes$^{4}$, 
\vspace*{0.3cm}
\\ {\LARGE 
A. Efstathiou$^1$,
A. Blain$^{5}$, 
M. Fox$^{1}$,
R. J. Ivison$^6$, 
T. Jenness$^{7}$, 
A. Lawrence$^{3}$,
\vspace*{0.3cm}
}\\ {\LARGE  
M. Longair$^{8}$,
S. Oliver$^{9}$, 
J.A. Peacock$^{3}$
\vspace*{0.1cm}
}\\
$^1$ Astrophysics Group, Blackett Laboratory, Imperial College, 
Prince Consort Road,London SW7 2BW, UK\\
$^2$ Centre for Astrophysics and Planetary Science, School of Physical
Sciences, 
University of Kent, Canterbury, Kent, CT2 7NZ, UK\\
$^3$ Institute for Astronomy, University of Edinburgh, Royal
Observatory, Blackford Hill, Edinburgh EH9 3HJ, UK\\
$^4$ Instituto Nacional de Astrofisica, Optical y Electronica (INAOE), 
Apartado Postal 51 y 216, 72000 Puebla, Pue., Mexico\\
$^5$ Institute of Astronomy, University of Cambridge, Madingley Road,
Cambridge CB3 0HA, UK\\
$^6$ UK ATC, Royal Observatory, Blackford Hill, Edinburgh EH9 3HJ\\
$^7$ Joint Astronomy Centre, 660 N. A'ohoku Place, Hilo, Hawaii 96720, 
USA\\
$^8$ Cavendish Astrophysics Group, Cavendish Laboratory, Madingley
Road, Cambridge CB3 0HE, UK\\
$^9$ Astronomy Centre, CPES, University of Sussex,
Falmer, Brighton BN1 9QJ, UK\\
}
\date{Accepted;
      Received;
      in original form 2001 May 22}
 
\pagerange{\pageref{firstpage}--\pageref{lastpage}}
\pubyear{2001}
\volume{}

\label{firstpage}

\maketitle


\begin{abstract} 
We present an extended analysis of the SCUBA 
observations of the Hubble Deep Field (HDF), 
expanding the areal coverage of the Hughes et al. 1998 study by a
factor of $\sim1.8$ and containing at least three further sources
in addition to the five in that study. 
We also announce the public release of the
reduced data products. The map is the deepest ever made in the
sub-millimetre, obtained in excellent conditions (median 
$850\mu$m optical depth of $0.16$).  
Two independent reductions were made, one with SURF and 
the other with a wholly algorithmic 
IDL analysis which we present in detail here.  
Of the three new sources, all appear to be 
at $z\stackrel{>}{_\sim}0.9$ 
and one is provisionally associated with 
an Extremely Red Object ($I-K>5$). 
%
%
%
%
%
%
There appears to be 
no significant cross-correlation signal between the $850\mu$m 
fluctuations and sources detected by ISOCAM, the VLA or Chandra, nor
with Very Red Objects ($I-K>4$), nor quasars and quasar candidates in
the HDF (notwithstanding a small  
number of individual weak candidate detections). This
is consistent with interpretations where the $850\mu$m-selected 
galaxies are at
higher redshifts than those currently probed 
by ISOCAM/VLA, and predominantly not Compton-thin
AGN. 
There are only one or two compelling cases for the radio source being
the sub-mm source. Nevertheless, most SCUBA-HDF point sources have a
nearby radio source apparently well-separated from the sub-mm
centroid. 
%
%
%
%
\end{abstract}

\maketitle

\section{Introduction}\label{sec:introduction}

Sub-millimetre blank-field surveys represent a major time 
investment on the JCMT
(e.g. Barger 
et al. 1999a,b, Lilly et al. 1999a,b, Eales et al. 1999, 2000, Fox et
al. 2002, Scott 
et al. 2002), and
will arguably be among the most 
important extragalactic surveys of the coming decade. 
Lensing cluster surveys (e.g. Smail, Ivison \& Blain 1997)
demonstrated early on the feasibility of sub-mm  
surveys with SCUBA (Holland et al. 1999) on the JCMT, and the
extremely deep integration in  
the HDF by ourselves (Hughes et al. 1998) showed that sub-mm surveys
are also feasible in blank fields. 
A total of $50$ hours were spent in the HDF, and
the central $90''$ radius portion of this data has already been
presented in Hughes et al. (1998); here we present the results
from the remainder of the map, extending substantially 
into the Hubble Flanking
Fields (HFF), and announce the public release of the reduced data products. 

This paper is structured as follows. In Section \ref{sec:method} we 
review the observing strategy and data acquisition; Section
\ref{sec:method2} discusses our data reduction algorithms and
considers the source astrometry and flux 
calibration uncertainties (Section 
\ref{sec:calibration}). Section
\ref{sec:results} presents the results of our reductions, discusses
the possibility of spatially correlated noise (Section
\ref{sec:flipped}), 
and presents a discussion of the source extraction 
(Section \ref{sec:deconvolution}). 
Details of the data products in public
release are presented in Section \ref{sec:release}. 
In section \ref{sec:associations} we cross-identify 
our point sources with HDF/HFF, ISO, Chandra and VLA sources, and
discuss the ambiguities with identifications of
sub-mm survey point sources. We also attempt to obtain statistical
sub-mm detections of sources not detected individually. 
Finally, in
Section \ref{sec:conclusions} we summarise our 
results.
%
%
%
%
%

\section{Data acquisition}\label{sec:method}
\begin{figure}
  \ForceWidth{3.0in}
   \BoxedEPSF{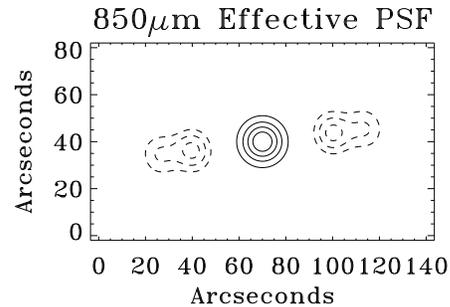}
\caption{\label{fig:psf}
Effective point spread function at $850\mu$m for the combined $30''$
and $45''$ chop throw data. Positive contours are in steps of $0.2$ of 
the maximum, starting at $0.2$. Negative contours (dashed) are in
steps of $-0.1$ starting at $-0.1$. North is up, and East to the
left. The effect of the chop tracking bug is included. 
}
\end{figure}

Details of the observations were presented in Hughes et al. (1998); we 
summarise the main points here. 
We observed the Hubble Deep Field (HDF)
with the Submillimetre Common User Bolometer
Array (SCUBA) at the James Clerk Maxwell Telescope. 
We used the $64$-point jiggle
pattern to 
obtain Nyquist sampled images at both $450\mu$m and $850\mu$m. 
We took a total of $88$ $\sim1$ hour integrations on the HDF. 
The
chop throw was set to $30''$ for $\sim48$ hours, and the remainder was 
spent using a chop throw of $45''$. 
This observational strategy was
intended as an experiment in minimising the sky fluctuation noise, but 
in retrospect it was fortunate since it gives powerful deconvolution
constraints (Hughes et al. 1998). 
There was a bug in the telescope
chop tracking software at the time, to the effect that the position
angle of the chop varied throughout each integration, though the
variation has since been quantified and its cause is well-understood. The 
resulting
effective point spread function for the combined chop throws is
plotted in Figure \ref{fig:psf}. 

Of the $130$ SCUBA-HDF demodulated data sets, only one (UT date
19980211 run $59$) was affected by the astrometric shift caused by the
clock error on the acquisition computer (Jenness 2000). 
This dataset has a pointing
shift of $24''$ (possibly indicating a rotation error rather than
clock shift), and has been excluded from the analysis below. The
maximum offset for the remaining $129$ 
demodulated data files is only $0.08''$. Of the $265$ calibration
observations, six  
have shifts $>1''$ (19980115 run $115$, and 19980116 runs $66-68$,
$75$, $76$). These were
excluded from the astrometric calibration below. 
The remaining calibrators all have shifts $<0.13''$. 

\begin{figure}
   \ForceWidth{3.0in}
   \BoxedEPSF{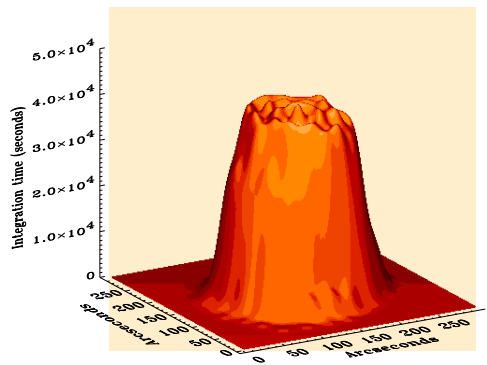}
\caption{\label{fig:mash}
Integration time per $14.5''$ FWHM beam at $850\mu$m for the
SCUBA-HDF. Note the steep sides and flat top. 
}
\end{figure}

The integration time per point decreases rapidly towards the periphery 
of the field: Figure \ref{fig:mash} shows the integration time as a
function of position for the long wavelength detector array. 

\section{Data analysis}\label{sec:method2}

\subsection{Reduction algorithms}

We made our reductions using SURF v1.2, the details of which are 
discussed in Hughes et al. (1998). In parallel we also reduced the 
data using  a 
specially-written pipeline in the commercial Interactive Data Language 
(IDL) package, discussed briefly in Hughes et al. (1998). 
In this subsection we describe the pipeline more fully. 

The IDL pipeline made a 
first-order fit to the  
two-dimensional sky background gradient 
before the combination of the nods. 
Combining the 
fit-subtracted nods gave a modest signal-to-noise improvement of 
$\sim5\%$. Bolometer astrometry was obtained using the {\sc 
scuba2mem} routine developed by T.Jenness. The factory flat field was
used, as in the SURF reduction. Extinction corrections for each
bolometer were determined 
according to the current best practice (Archibald, Wagg \& Jenness
2000) of using the smoothed CSO $225$ GHz optical depths where
available in conjunction 
with canonical conversions, or $850\mu$m skydips to estimate the
optical depths at $850\mu$m and $450\mu$m where the $225$ GHz data are
not available. 
This method of extinction correction has become standard practice in
later 
versions of SURF but was not used in the Hughes et al. (1998)
analysis. 
We used a $6$th order polynomial fit to the archival $225$ GHz data to
model the short timescale variations, but used the skydips where less
than $7$ reliable CSO measurements are available within $\pm 1$ hour
of each observation. 

Further differences between the SURF and IDL reduction algorithms are
in the bolometer deglitching and sky subtraction. In the SURF reduction,
noisy bolometers were identified and eliminated by hand, and an
instantaneous sky
level to be subtracted from the beamswitched data was determined from
either a median level or from 
interactively-chosen bolometers. SURF also has the facility for
timeline despiking. 
In the IDL pipeline, the approach
was instead to iterate on the following procedure: (a) make noise
estimates for each bolometer in 
$128$-readout groups from Gaussian fits to the readout histograms; (b) 
perform an N$-\sigma$ clip deglitching using these time-dependent
noise estimates; (c) determine and subtract a zero-point modal sky level
in each readout from a fit to the readout histograms of all bolometers 
in the array. In other words, each iteration makes a temporal noise
estimate and deglitching, followed by a spatial sky subtraction. 
Each iteration performs successively harder deglitching
cuts. Noisy bolometers are not eliminated by hand, but are instead
automatically assigned a low inverse variance weight. 
This algorithm is not affected by the presence of sources,
since there are no sources in the HDF field bright enough to be 
detected significantly in any single jigglemap, let alone readout. 
However, the same is not true of the calibrators. For finding glitches 
and determining noise levels in these,
a timeline without object signal was constructed by subtracting the
mean of the immediately previous and subsequent readouts from each
readout. 
Our noise estimation is very different in philosophy and practice to
(e.g.) the Canada-UK survey (e.g. Eales et al. 1999, 2000). 

The final maps were constructed using an
optimal noise-weighted drizzling 
algorithm (see Fruchter \& Hook 1997), with the footprint set to 1
square arcsecond.  
(SURF also offers noise weighted mosaicing in the REBIN
mosaicing routine, developed for the Hughes et al. 1998 HDF analysis,
though the noise is estimated by a different algorithm.)
In any square-arcsecond pixel, the flux is the noise-weighted average
of the bolometer readouts at that position. As discussed above, the
noise for each bolometer readout had been estimated from Gaussian fits 
to the readout histograms. The noise map was constructed in a similar
fashion, with each pixel containing the formal error on the noise
weighted average (i.e. $\sqrt{\Sigma(1/\sigma^2)}$). By the central
limit theorum the noise statistics in our final map (as indeed in the 
REBINned equivalent) should be approximately Gaussian. 

The $1''$ footprint maps are an attempt to represent the detectors'
views of the sky 
at each position. However the signal-to-noise per pixel is
very low because the bolometers have been interleaved as far as
possible rather than
coadded, and no smoothing has yet been applied. 
For example, convolution with the point spread function is a very
common method of 
extracting point sources (e.g. Eales et al. 1999, 2000 in the context
of extragalactic sub-mm surveys) and is an optimal 
point source filter in the case of uniform noise. Here we discuss the
non-uniform noise case. 

To search for sources in the $450\mu$m maps, we minimised
the chi-squared of a 
Gaussian PSF fit with the instrumental FWHM  
($7''$ at $450\mu$m) at every position. This can be expressed as a
convolution: if $S$ is the image signal, $P$ the PSF 
and $W$ the reciprocal of the image variance, then the best fit flux F
is given by: 
\begin{equation}\label{eqn:convol}
F = \frac{(SW)\otimes P}{W\otimes P^2}
\end{equation}
where $\otimes$ denotes a convolution.                 
The error on this best-fit flux is given by: 
\begin{equation}\label{eqn:convol_err}
(\Delta F)^2 = \frac{1}{W\otimes P^2}
\end{equation}
(These relations are derived in the Appendix.)
Setting a threshold in the $F/\Delta F$ map is formally optimal
as a source extraction algorithm. Note that the $F/\Delta F$ map is
not at all optimal for spatial resolution -- in fact we have gained as
much point source sensitivity as possible at the expense of spatial
resolution in creating this map. Nevertheless, the centroids of the
point sources are as 
accurately determined as possible in these maps. Since our sources are 
not expected to be resolved or 
confused
at $450\mu$m, 
the loss of spatial resolution is of no consequence. 
This algorithm has also been applied to our ongoing wide-area $850\mu$m
$8$ mJy survey (Scott et al. 2002). 

Extraction of sources in the $850\mu$m map is more difficult, as the
noise-weighted PSF convolution is only optimal if there are typically
$\ll 1$ source per beam. This does not hold for the centre of the
$850\mu$m map (though the relative noise contribution of source
confusion drops strongly outside the area considered by Hughes et
al. 1998). Instead we place model point sources at
observed peaks, and by simultaneously varying their fluxes we minimise
the total $\chi^2$ of the map. Positions for these peaks were
determined from the peaks in a noise-weighted convolution, though we
also tried using the Hughes et al. (1998) CLEANed positions. 



\subsection{Calibration}\label{sec:calibration}


In a beam map of Uranus taken during the run, the positions in the
short and long arrays 
agreed to better than an arcsecond. CRL$618$, IRC$+10216$ and
OH$231.8$ typically showed a slight ($\sim 2''$)
shift between $450\mu$m and $850\mu$m positions, assumed intrinsic
to the source. 
The $850\mu$m position of CRL618 
agrees with the corresponding NVSS position (accurate to only 
$5.7''\times3.8''$, Condon \& Kaplan 1998), 
though the $450\mu$m position lies just outside 
the $1\sigma$ VLA error box. The JCMT pointing grid is believed
accurate to around $1''$. The 
IRAM $1.3$ mm interferometric position determined by
ourselves 
and others (Downes et al. 1999) is offset from our position of the
brightest $850\mu$m source by 
$2.5''$. This is a roughly $3\sigma$ discrepancy with the position
published in Hughes et al. 1998 if we use $\pm\theta_{\rm
FWHM}/(2\times S/N)$ as the astrometric  
uncertainty (where $\theta_{\rm FWHM}$ is the beam FWHM and $S$, $N$
are the signal and noise respectively), but it may be
made consistent if the noise $N$ includes a roughly equal confusion
noise term added in quadrature (Hogg 2000), 
and/or a systematic error of order
$1''$ in the JCMT pointing grid solution. Such a systematic could
exist at the $\sim$ arcsecond level in some parts of the
azimuth--elevation plane and be consistent with 
existing inclinometry and 
pointing checks. 


Flux calibration was obtained from planets and from CRL618, OH$231.8$
and IRC$+10216$. Using the peak flux of point sources in the $1''$
drizzling footprint IDL maps
yields flux conversion factors of $223\pm15$ Jy/V at $850\mu$m and
$618\pm 115$ Jy/V at $450\mu$m, where the errors have been estimated
from the variance among the calibrators. These values are consistent
with the advertised conversions valid at the time for the narrow-band
filters. 

\section{Results}\label{sec:results}

\begin{figure*}
   \ForceWidth{3.0in}
   \hSlide{1.0cm}
   \BoxedEPSF{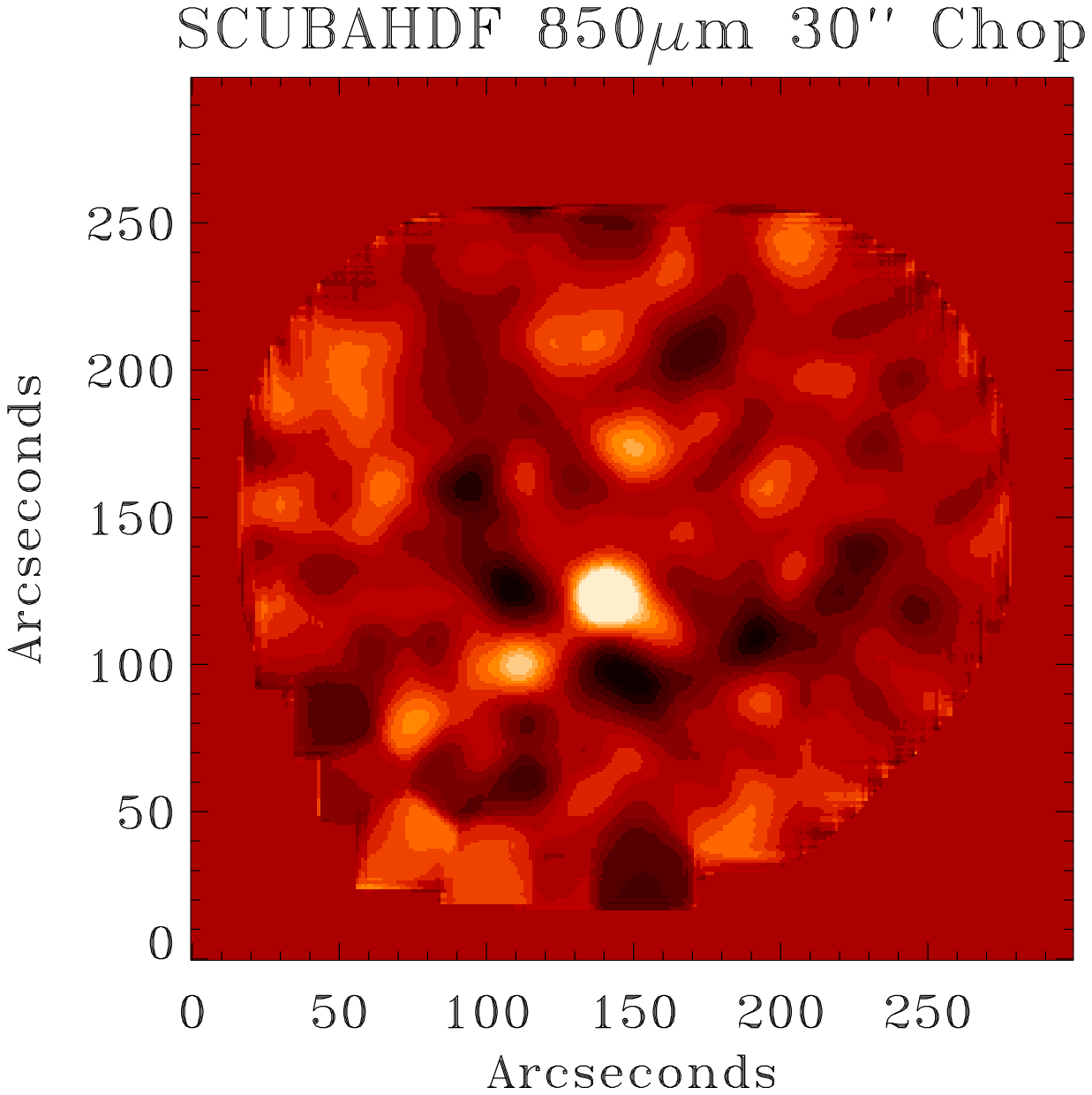}
   \ForceWidth{3.0in}
   \hSlide{-1.0cm}
   \BoxedEPSF{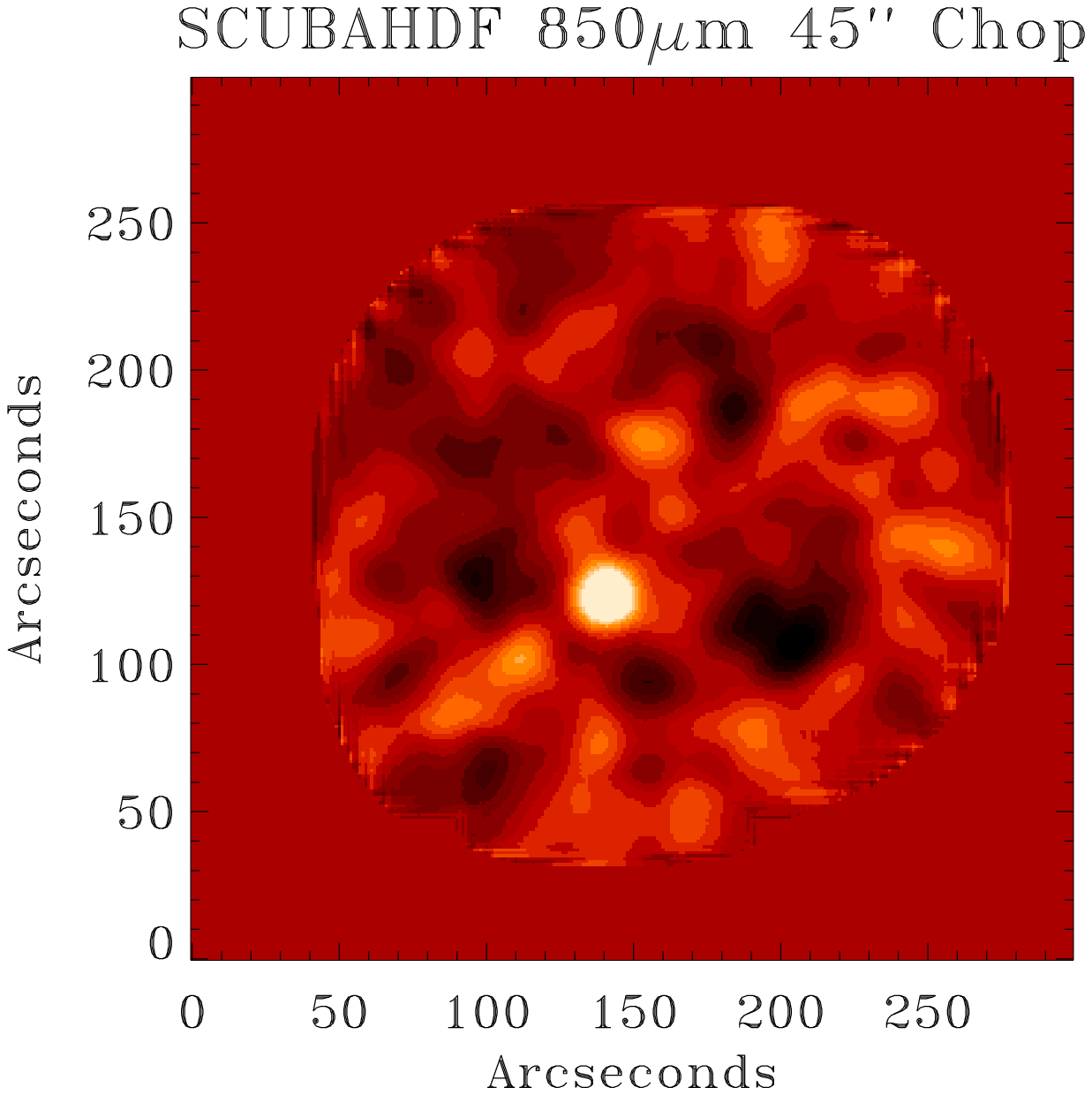}
 \vspace*{0.5cm}
   \ForceWidth{3.0in}
   \hSlide{1.0cm}
   \BoxedEPSF{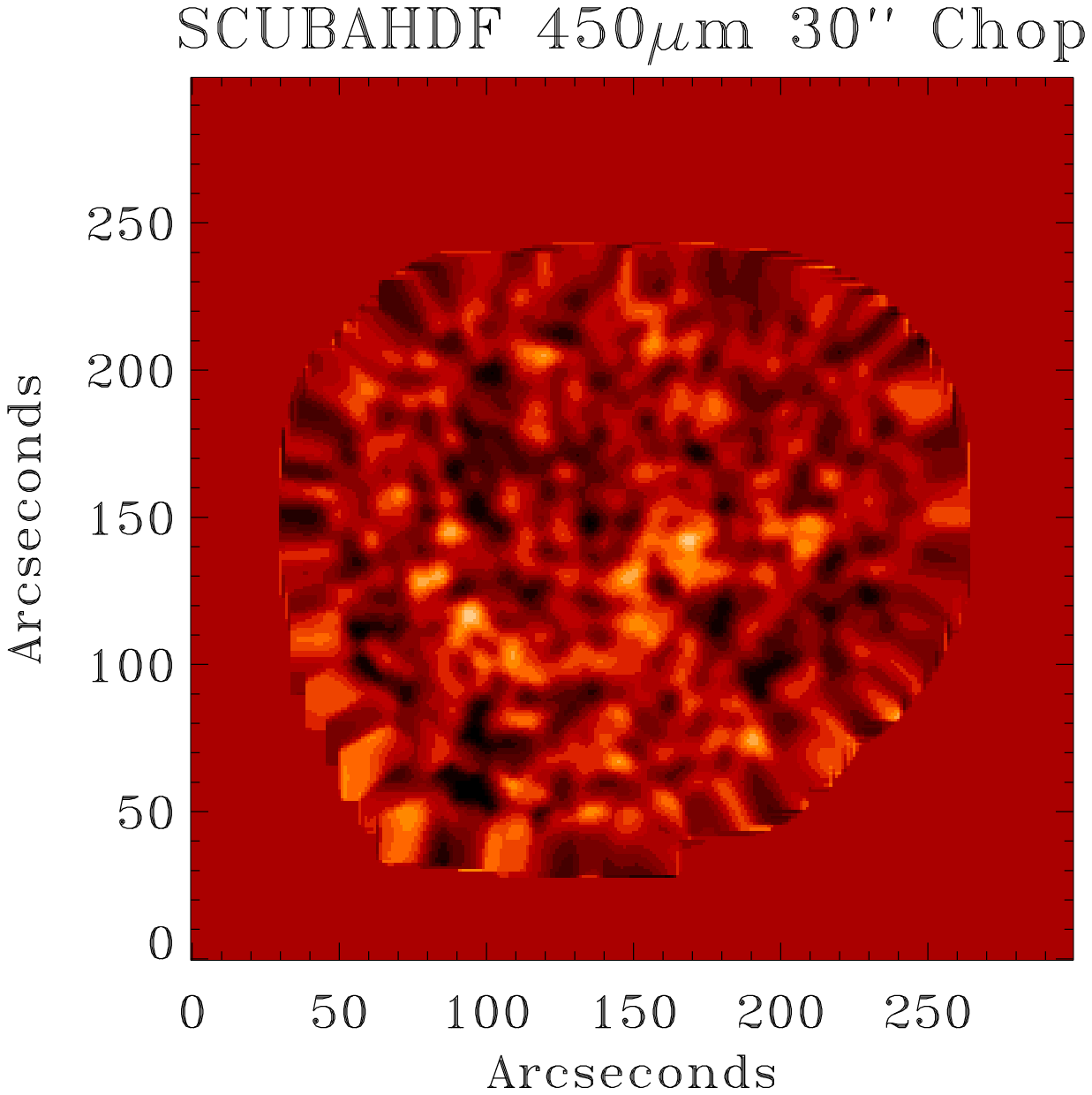}
   \ForceWidth{3.0in}
   \hSlide{-1.0cm}
   \BoxedEPSF{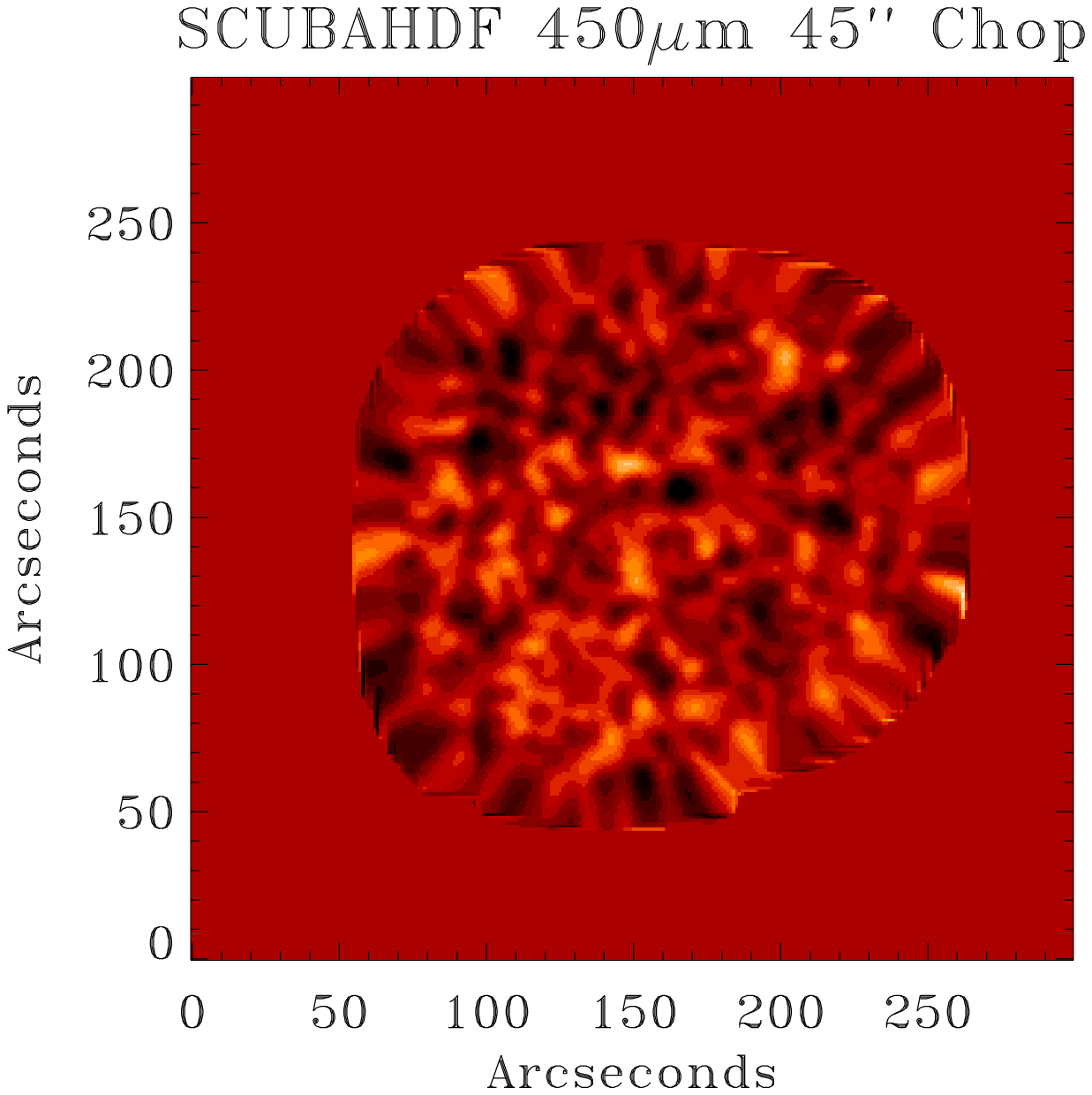}
\caption{\label{fig:raw_images}
SCUBA images of the Hubble Deep Field prior to deconvolution. The
$1''$ drizzling footprint images have been convolved with the positive
part of the instrumental 
point spread function, using the noise-weighted convolutions given in
equations \ref{eqn:convol} and \ref{eqn:convol_err} with $P$ set to
a Gaussian with FWHM of $14.5''$ at $850\mu$m and $7''$ at
$450\mu$m. Note that the 
maps are not fully-sampled at the edges. North is up, East to the
left, and white indicates high flux values. 
}
\end{figure*}

\begin{figure*}
 \vspace*{-11cm}
   \ForceWidth{5.0in}
   \vSlide{2cm}
   \BoxedEPSF{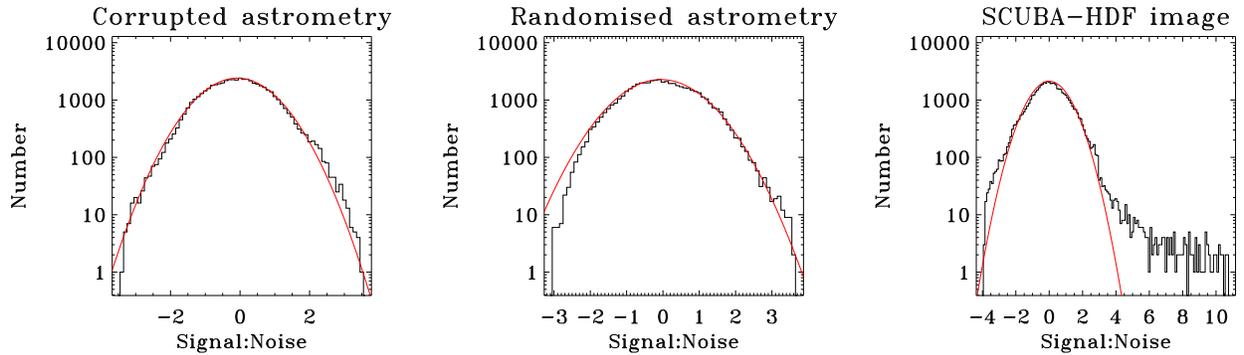}
\caption{\label{fig:pd_histos}
Fluctuation histograms for the $30''$ chop throw 
corrupted map (left), randomised map
(centre), and the original  $30''$ chop SCUBA-HDF map, all after a
noise-weighted convolution with a $7.25''$ FWHM Gaussian (half the
width of the JCMT primary beam). 
Iterative Gaussian fits were made to the histograms, rejecting
outlying points, and are overplotted on the Figure. The means are
consistent with zero and the variances with unity, as expected.
}
\end{figure*}

\subsection{Spatially correlated noise}\label{sec:flipped}
The raw images with noise-weighted PSF convolutions are shown in
Figure \ref{fig:raw_images}. 
The minimum noise level in the combined $30''$ and $45''$ maps is
$0.39$ mJy per beam, using a $14.5''$ FWHM Gaussian beam. Using the
full beam profile (i.e., including the negative sidelobes) improves
this by approximately $\sqrt{2/3}$. 
It is possible that some of the
fluctuations in these 
images are spatially correlated, due to e.g. unsubtracted 
structure in the (terrestrial) sky background. In order to quantify
the level of this effect, we made a further parallel IDL reduction
with the 
astrometry of the jiggle maps intentionally corrupted with arbitrary
$n\pi/3$ rotations and/or reflections, and the offsets of the jiggle
pattern 
reversed. Zero rotations were excluded. 
In mosaicing the jigglemaps to a final coadded image, a
further arbitrary $n\pi/2$ reflection and/or rotation was added. 
Photometry observations were excluded from these mosaics, since the
photometry targets lie on the
invariant point of the rotations -- i.e., the photometry
observations are designed to have a bright source at the centre, so
rotating the jigglemap around its centre would not smear out the
source. 
These transformations will have the effect of smearing
out genuine extragalactic sky structure, while preserving the noise
level due to short-term 
terrestrial sky fluctuations. These corrupted maps therefore place an
upper limit on the level of spatially correlated noise. 
No point sources were evident in these corrupted maps to a level of
$4\sigma$. We measured the $1\sigma$ fluctuations by iterative
Gaussian fits to the main part of the maps, and found the flipped maps 
to have almost exactly the same noise level ($\sim 1\%$ higher in the
flipped map). 
As a further 
comparison, randomising the individual
bolometer astrometry yields an estimate of the spatially 
{\it uncorrelated} component, which gave a $1\sigma$ noise $\sim 1\%$
lower than the uncorrupted data, implying the noise is spatially
uncorrelated to high precision. 
The fluctuation histograms for the $30''$ chop throw maps are shown in 
Figure \ref{fig:pd_histos}; note the Gaussian nature of the corrupted
maps. 

\subsection{Cirrus noise}\label{sec:cirrus} 
How much of the fluctuations in the SCUBA-HDF maps (Figure
\ref{fig:raw_images}) are due 
to cirrus noise? To estimate this, we need to know the power spectrum 
of cirrus fluctuations on scales down to the beam size. The power 
spectrum is observed to have a steep $k^{-3}$ slope  
from the largest spatial scales probed by DIRBE (Wright 1998) to  
arcminute scales probed by IRAS and ISOPHOT (Gautier et al. 1992, 
Herbmeister et al. 1998, Lagache \& Puget 2000). On  
sub-arcminute scales, the cirrus structure traced by extinction 
measured by ISOCAM (Abergel et al. 1999) continues the $k^{-3}$ 
dependence to $\sim 6''$ scales, so that no characteristic scale has 
yet been detected for cirrus fluctuations.  

Using the models of Gautier et al. (1992) with a $k^{-3}$ power-law 
power spectrum at all scales, we obtain the following 
expression for the RMS $850\mu$m cirrus fluctuations with a $45''$ 
chop throw: 
\begin{equation} 
\sigma_{850\mu{\rm m}} (\mu{\rm Jy}) \simeq 5 I_{100}^{3/2} 
\end{equation} 
where the cirrus $100\mu$m background level $I_{100}$ is given in 
MJy/sr,  
$\sigma_{850\mu{\rm m}}$ is in $\mu$Jy, 
and we have assumed $I_{100}/I_{850}=14.2$ (Lagache et al 1999).  
Using $k^{-2.6}$ or $k^{-3.8}$ power spectra (the range spanned  
by the Gautier et al. models) modifies the numerical coefficient to 
approximately  
$11$ or $1$ respectively. Using a chop throw of $30''$ reduces the 
cirrus noise by $\sim30-60\%$. 
Together with the $\times 2$ uncertainty in the
power spectrum
normalisation (corresponding to $\sqrt{2}$ uncertainty in the RMS
noise), this 
expression should be correct to around an order of 
magnitude. 

The IRAS
Sky Survey Atlas (ISSA) gives an estimate of the $I_{100}$ background
in the HDF 
area. This obtains $0.25\pm0.07$ MJy/sr in a $30'\times30'$ area
centred on the HDF, consistent with the ($4'$ resolution) ISSA
measurement at the HDF position itself of $0.331$ MJy/sr. Neither 
value
has been corrected for extragalactic background light
contribution. These values are nevertheless
slightly lower than the Schlegel et
al. (1998) estimate of $0.61$ MJy/sr. 

In either case, the cirrus fluctuations at $850\mu$m are 
negligible on the 
scale of the beam, and it can be similarly shown that the same applies 
at all sub-mm wavelengths for all current and future planned
extragalactic sub-mm surveys on the beam scales, and on all scales in
our SCUBA-HDF map according to the Gautier et al. models 
(e.g. Blain 1999). 

\subsection{Source extraction}\label{sec:deconvolution}

\subsubsection{Extraction algorithm}

The $850\mu$m map is close to the formal confusion 
limit (one source per 25-40 beams), 
so we cannot 
extract sources by simply convolving with the beam. 
Instead, in Hughes 
et al. (1998) we
iteratively extracted sources using the CLEAN algorithm. Using
numerical simulations we found that point sources were distinguishable 
from blends and confusion noise peaks by the fact that they appear at
around the same position in both chop throw maps. Thus the number of
claimed point sources in Hughes et al. 1998 is much lower than the
number of distinct peaks found in the map. (Of course, this is not to
say that none of the remaining peaks are real.) 

\begin{table*}
\begin{tabular}{llllll}
  Name   & \multicolumn{2}{l}{Position (J2000)}  & $S_{850}$ (mJy) & S/N  at           & $S_{450}$ (mJy)  \\
          &  &   &                & $850\mu$m    &   \\
  
~HDF850.1   &      12 36 52.22 & +62 12 26.5 & 5.6 $\pm$  0.4      & 15.3     & ~2.1 $\pm$  4.1\\
~HDF850.2   &      12 36 56.50 & +62 12 03.5 & 3.5 $\pm$  0.5      & 7.6      & \hspace*{-0.05cm}14.1 $\pm$  5.5 \\
(HDF850.3)  &      12 36 44.35 & +62 13 07.5 & 1.0 $\pm$  0.5      & 2.1      & ~5.9 $\pm$  4.3\\
~HDF850.4   &      12 36 50.37 & +62 13 15.9 & 1.1 $\pm$  0.2$^*$  & 
& ~1.3 $\pm$  4.0\\
~HDF850.5   &      12 36 51.98 & +62 13 19.2 & 1.0 $\pm$  0.2$^*$  &   & ~0.6 $\pm$  4.0\\
 \\
~HDF850.6  &      12 37 01.21 & +62 11 45.3 & 6.4 $\pm$  1.7      & 3.8      & \hspace*{-0.15cm}$-$24 ~$\pm$  54\\ 
~HDF850.7   &      12 36 35.20 & +62 12 42.4 & 5.5 $\pm$  1.5      & 3.7      &  \hspace*{0.1cm}~~~N/A\\
~HDF850.8   &      12 36 53.07 & +62 13 54.5 & 1.7 $\pm$  0.5      & 3.5      & \hspace*{-0.135cm}$-$5.8 $\pm$  5.7\\
 \\
\end{tabular}
\vspace*{-2.5cm}
\begin{tabular}{lll}
~~~~~~~~~~~~~~~~~~~~~~~~~~~~~~~~~~~~~~~~~~~~~~~~~~~~~~~~~~~~~~~~  & \mbox{\Huge \} \normalsize\hspace*{-0.25cm} 5.1$^*$} & ~~~~~~~~~~~~~~\\
\end{tabular}
\vspace*{2.5cm}


\caption{\label{tab:sourcelist}Sources from the $850\mu$m map of the
Hubble Deep Field. The quoted $450\mu$m fluxes are the values at the quoted positions
in the noise-weighted, beam-convolved $450\mu$m map. 
HDF850.3 is included for consistency with the Hughes et al. (1998) paper, 
though it formally fails our revised source extraction algorithm. 
$*=$ These two sources were detected 
as a single, blended source in the extraction algorithm. We have adopted
the deconvolution of Hughes et al. (1998) to deblend these sources, and
divided the total flux 
between the two according to the Hughes et al. ratio. The quoted 
signal-to-noise is for the combined source. HDF850.7 lies on the edge of 
the $850\mu$m map and is not covered by the $450\mu$m image. 
}
\end{table*}

Our approach here differs from the Hughes et al. (1998) analysis. 
Using the noise-weighted convolution peaks as a starting point, we
make a simultaneous fit to both the chop throw maps. 
We simultaneously vary the fluxes (but not positions) 
of the point sources and obtain a
solution minimising the total $\chi^2$. 
This procedure also naturally yields a signal-to-noise for each
recovered source. This minimisation is discussed in more detail in
Scott et al. (2002). Flux boosting caused by point sources being
blended with neighbouring sources (i.e. boosting connected to
confusion noise) is a possibility in our map (e.g. 
Eales et al. 2000; Scott et al.2002; Borys et al. 2001), 
but one which depends on
the clustering properties of the sources in question (e.g. Peacock et
al. 2000). To avoid the source list being overly dependent on
simulation assumptions we quote only fluxes uncorrected by boosting 
in this paper. 

\subsubsection{Comparison with Hughes et al. 1998}
The results of
this analysis are in general in excellent agreement with the analysis
presented in Hughes et al. (1998), though certain subtle differences
are worth noting. Firstly, our flux conversion factor is slightly
different to that used in Hughes et al. (1998), resulting in a
slightly lower flux for the brightest source. Secondly, the close pair 
of sources HDF850.4 and HDF850.5 were not deblended in this algorithm, 
although the combined ``source'' is clearly extended. Our CLEAN
algorithm is more effective at deblending such pairs, so we adopt the
flux ratio in Hughes et al. 1998 for these two sources. The $\chi^2$
map using the sources at the CLEAN-deconvolved positions does not show 
obvious areas of poor fit. However, we
caution that the map shows hints of a more complicated structure here. 
Thirdly,
HDF850.3 appears rather fainter in the $45''$ chop throw map, so that
it would fail the strict confirmation criterion used in Hughes et
al. 1998. This particular source flux appears to be rather sensitive
to the algorithm used to mosaic the map. The 1998 IDL analysis used a
circular drizzling footprint of the same size as the bolometers, and
the SURF map used the SURF REBIN task. The source appears in these to
lie on a ridge of extended emission in these maps, 
which pushes it over the
threshold. However in the current reduction using the $1''$ drizzling
footprint, convolved with the beam, the ridge is less
pronounced and the source drops below the threshold, to $2.1\sigma$.
The fact that the 
mosaicing and not the reduction is the key difference is confirmed by
mosaicing the original 1998 
SURF reduction with the current $1''$ drizzling
footprint method: we again find the source below the threshold. In
Table \ref{tab:sourcelist} we list the confirmed point sources found
in both maps, with HDF850.3 in brackets for this reason.

\subsubsection{Reliability and completeness}\label{sec:reliability}
We tested our reliability and completeness by performing the source 
extraction on simulated maps. False positives were defined as sources
extracted from the maps where no single simulated source within the beam
contributes
$>50\%$ of the flux. 
Simulated 
sources were treated as missing if there was no extracted source within
the beam with a flux within a factor of $2$ of the simulated flux. 
We tried our source extraction using peaks from the maps convolved
with a $14.5''$ beam, a $7.5''$ beam and a $5''$ beam. The results are
shown in Figure 5. 
In this figure, sources are considered genuine
if any single simulated source within the beam contributes $>50\%$ of
the observed flux. The reliability is defined as the fraction of
extracted sources which pass this criteria for being genuine.

The results indicate that basing our fits on the peaks in the $14.5''$ 
map give the highest completeness and reliability, but unless
the signal:noise cut is very high both only reach $\sim 80\%$. 
This reliability implies that of the $7$ sources we extracted by 
this method, we expect 
$1.4$ on average to be a blend. This agrees well with the fact 
that HDF850.4 and HDF850.5 from Hughes et al. (1998) are extracted
as a single source, albeit extended. We adopt the CLEAN deconvolution 
from Hughes et al. (1998) to divide the flux for this source between
the two. Nevertheless, we cannot exclude the possibility that our
source list is entirely free of further blends. 

No negative point sources were detected that were not
associated with sidelobes of the positive sources. 


\begin{figure}
   \ForceWidth{3.5in}
   \hSlide{-0.5cm}
   \BoxedEPSF{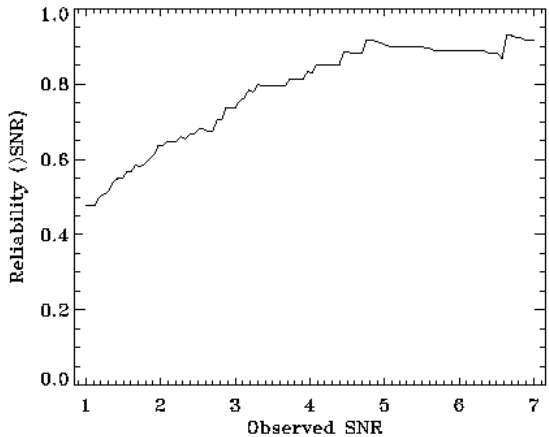}
\label{fig:reliability_plot}
\caption{
Reliability for the source extraction algorithm. 
Extracted sources were considered genuine if a simulated
source lay within the beam and contributed at least 
$50\%$ of the observed flux. 
}
\end{figure}

\subsubsection{Additional sources and structure in the map}

The point sources selected for this $\chi^2$ fitting were selected 
from the area covered by both the $30''$ and $45''$ chop data, because
a simultaneous fit to both maps is intrinsic to our algorithm. 
However, one further source (HDF850.6) 
would pass the source extraction threshold
if we require fitting in either map on its own. 
This source is just outside the area covered by the
$45''$ chop throw map and no useful limits can be obtained at its
position from either $450\mu$m map. The $30''$ chop throw map is only
just fully-sampled at this position. 
The source
is on the periphery of the map where the effects of confusion are
much less dominant, and the source is also identified in Section 5 with a
VLA source and an extremely red galaxy. 
The VLA position is $2.8''$ from the SCUBA centroid, and the shift is in 
the opposite 
sense to the shift between the Downes et al. (1999) IRAM position of
HDF850.1 and the Hughes et al. (1998) SCUBA position of the same
source.
We therefore include this
source in  Table  \ref{tab:sourcelist}.
%
%

There is plenty of 
evidence for point sources outside the area of the map. For 
example, there is a $\sim3\sigma$ feature in the vicinity of $12 36
44 +62 14 20$ in both chop throw maps, although the best-fit
centroid is several arcseconds to the north off the edge of both
maps. There is also an apparent trough of negative emission to the SW
of the brightest source, seen in both chop throws. We carefully
examined the individual jigglemaps to test whether this is due to an
incorrectly-weighted noisy bolometer precessing around the jigglemap
centre as each night's observations progressed. We found no evidence
for such an effect in the individual jigglemaps (all of which appear
to have Gaussian signal:noise histograms with unit variance); the
apparent trough only appears in the coadded maps. This effect cannot 
be a Sunyaev-Zel'dovich decrement ($\lambda<1.4$mm) 
but can be attributed to
confusion noise from the unresolved point sources (recall the negative 
sidelobes of the beam). This serves to highlight the fact that our map 
contains much more information than the point sources extracted
above. 

\begin{figure*}
  \ForceWidth{5.0in}
   \hSlide{1.0cm}
   \BoxedEPSF{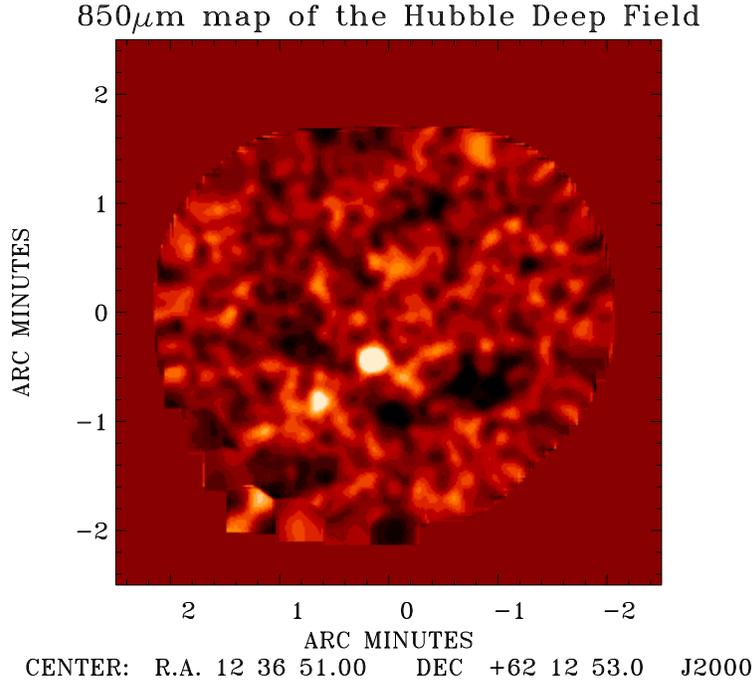}
\caption{\label{fig:scubahdf}Deconvolved $850\mu$m signal-to-noise 
map of the Hubble
Deep Field. North is up, and East to the left. Note that the edges of
the map are not fully-sampled.}
\end{figure*}

In Figure \ref{fig:scubahdf} we show what amounts to a 
CLEAN reconstruction of our $850\mu$m
map. The point sources in Table 
\ref{tab:sourcelist} were first subtracted from each chop throw map 
using the theoretical PSF incorporating the SCUBA chop tracking
bug. We coadded the $30''$ and $45''$ chop residual maps, and then
convolved this combined residual 
map with a $7.5''$ FWHM Gaussian beam. Finally, we added 
$14.5''$ FWHM Gaussian beams to the convolved residual map
at the positions of these sources, with
$\sqrt{3/2}\times$ the peak flux of the original point sources, which
approximates to the S:N gain in using the sidelobes. 
Thus, we effectively turn an [$-0.5,1,-0.5$] beam into simply
[$\sqrt{3/2}$]. 

\subsubsection{$450\mu$m sources}
No sources were detected in the $450\mu$m maps to a typical
$4\sigma$ per beam 
depth of around $15$ mJy, using a search for peaks in the noise-weighted
convolved images. The $450\mu$m fluxes at the positions of our
$850\mu$m sources are listed in Table \ref{tab:sourcelist}. 
If the hint
of a $450\mu$m detection of HDF850.2 is real, this would place it
at $(1+z)/(T/40K) = 2.0-5.1$. 




\subsection{Data release}\label{sec:release}
We announce the public release of the reduced data to the community. 
These are available at {\tt
http://astro.ic.ac.uk/elais/scuba\_public.html}, or from the authors. 
As well as the source lists quoted in this paper, the signal, noise
and signal-to-noise maps at both wavelengths are available, for each
chop throw separately and combined. Noise weighted point source
convolutions are available for a variety of PSF widths. 
In addition, the parallel reductions with
intentionally corrupted astrometry which quantify the level of
spatially correlated noise, are also released as an IDL save set. 
We are also releasing the 
SURF-reduced products from our Hughes et al. (1998) analysis. 

%
%
%
%

\section{Associations}\label{sec:associations}

In this Section we discuss the association of the 8 sources in Table
1 with objects detected in surveys of the HDF/HFF region undertaken 
in other wavebands; namely the optical surveys of Williams et al. (1996)
and Barger et al. (1999c), the radio survey of Richards (2000) and the
X--ray survey of Brandt et al. (2001).

\begin{figure*}
\epsfig {file=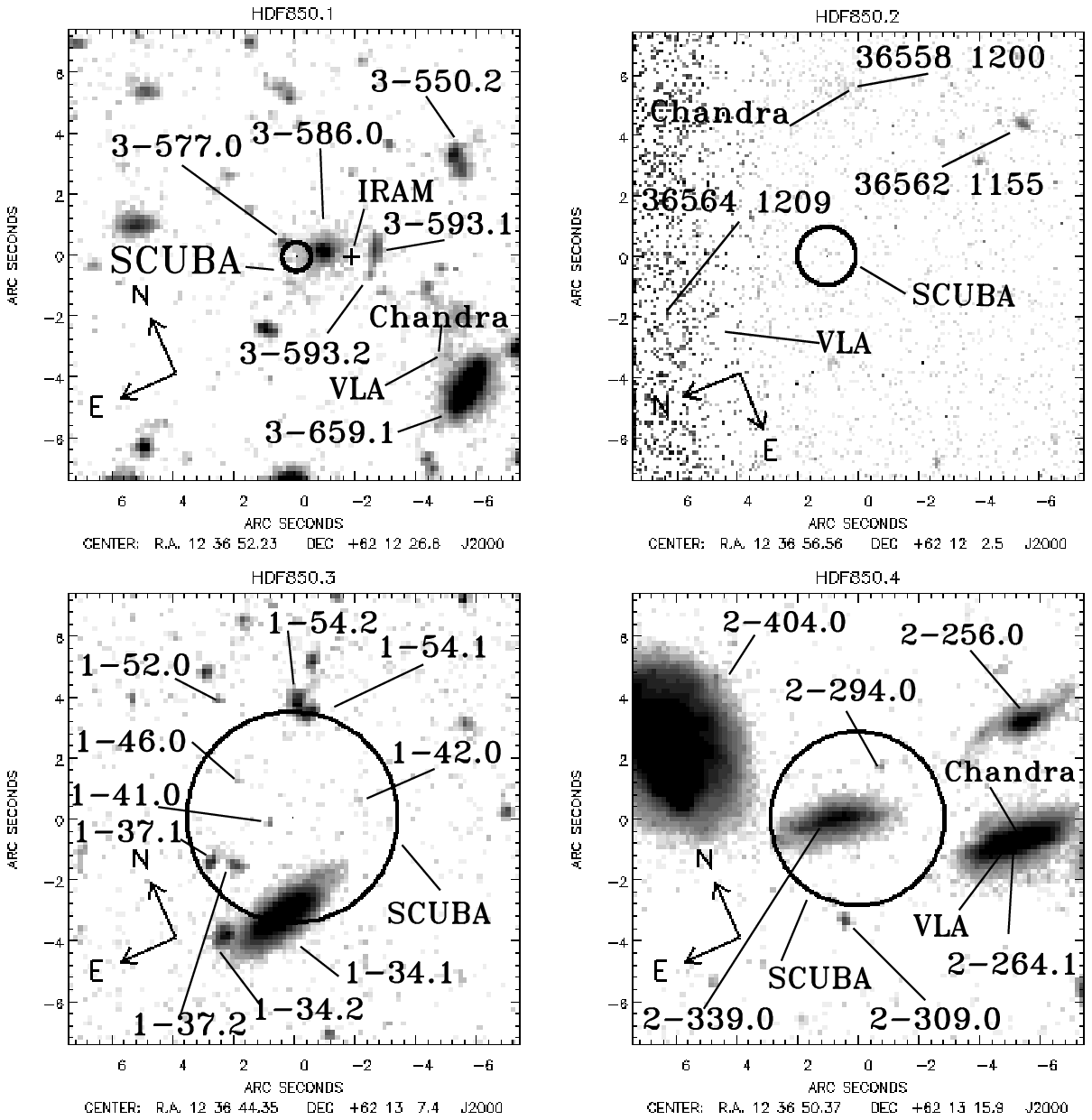,width=16cm,angle=0}
\caption{$I_{814}$ band postage stamp images for the eight SCUBA sources 
from Table 1. The circle in the centre of each image is the nominal 1$\sigma$
error circle for the SCUBA source, with radius equal to $\theta_{\rm FWHM}/
2 (S/N)$, where  $\theta_{\rm FWHM}$=14.7 arcsec is the FWHM of the beam
at 850$\mu$m and $S/N$ is the signal--to--noise ratio of the source detection,
taken from Table 1. (Note that we show in Section
\ref{sec:associations_scuba_vla} 
that confusion leads this to 
underestimate the true astrometric uncertainties.) 
Also marked are the names of optical galaxies (from the
catalogue of Williams et al. 1996 for HDF850.1, HDF850.3, HDF850.4, HDF850.5, 
and HDF850.8, and that of Barger et al. 1999c for HDF850.2, HDF850.6 and 
HDF850.7) and the locations of VLA sources from Richards
(2000) and Chandra sources from Brandt et al. (2001).}
\label{fig:postage}

\end{figure*}

\begin{figure*}
\epsfig {file=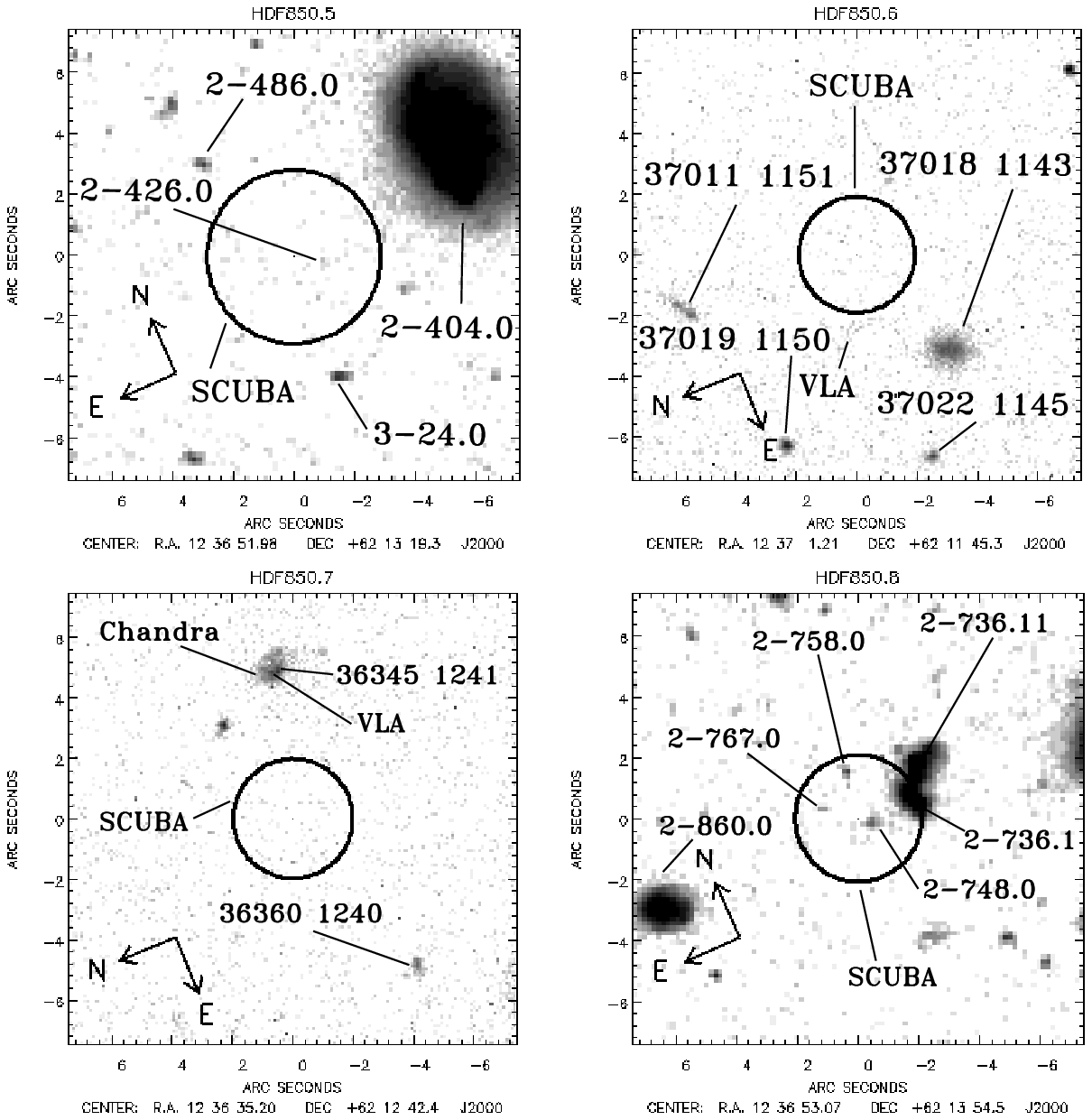,width=16cm,angle=0}
\contcaption{$I_{814}$ band postage stamp images for the eight SCUBA sources 
from Table 1. The circle in the centre of each image is the nominal 1$\sigma$
error circle for the SCUBA source, with radius equal to $\theta_{\rm FWHM}/
2 (S/N)$, where  $\theta_{\rm FWHM}$=14.7 arcsec is the FWHM of the beam
at 850$\mu$m and $S/N$ is the signal--to--noise ratio of the source detection,
taken from Table 1. Also marked are the names of optical galaxies (from the
catalogue of Williams et al. 1996 for HDF850.1, HDF850.3,HDF850.4, HDF850.5, 
and HDF850.8, and that of Barger et al. 1999c for HDF850.2, HDF850.6 and 
HDF850.7) and the locations of VLA sources from Richards
(2000) and Chandra sources from Brandt et al. (2001).}
\end{figure*}

In Figure~\ref{fig:postage} we present postage stamp images of the 8 SCUBA
sources from Table 1. At the centre of each postage stamp we overlay, on an 
$I_{814}$ band HDF or HFF image from Williams et al. (1996), the nominal 
1$\sigma$ positional error circle for the SCUBA source, with a radius given
by the conventional formula of $\theta_{\rm FWHM}/ 2 (S/N)$, where  
$\theta_{\rm FWHM}$=14.7 arcsec is the FWHM of the beam at 850$\mu$m and 
$S/N$ is the signal--to--noise ratio of the source detection, taken from Table
1. We note the names of nearby optical objects, from the catalogue of 
Williams et al. (1996) for the five sources (HDF850.1, HDF850.3,HDF850.4, 
HDF850.5, 
and HDF850.8) within the HDF and from that of Barger et al. (1999c) for the
three HFF sources (HDF850.2, HDF850.6 and HDF850.7), and also mark the 
positions of VLA sources from Richards
(2000) and Chandra sources from Brandt et al. (2001).

\subsection{Possible SCUBA--VLA associations}\label{sec:associations_scuba_vla}

One striking feature of this Figure is that five out of the eight sources
lie within $\sim$6 arcsec of a VLA source (three of which are also
included in the deep Chandra catalogue of Brandt et al. 2001) which is a highly unlikely 
chance occurence, given the low surface density of VLA objects on the sky.
To quantify this,  we may compute the probability
$P_0=1-\exp(-\pi N d^2)$ that the nearest object drawn from an unclustered 
population of surface density $N$ (such as VLA objects at least as bright as 
the putative radio counterpart of the SCUBA source) should lie no further
than $d$ from the SCUBA source position under the assumption that the SCUBA
and VLA source populations are uncorrelated. For the potential SCUBA--VLA
associations with this set of five  
sources (HDF850.1, HDF850.2, HDF850.4, HDF850.6 and HDF850.7) the $P_0$ values
so computed are 0.05, 0.03, 0.04, 0.01 and 0.01, respectively.

Similarly, Richards (1999) noted that as many as four of the
five SCUBA--HDF sources from Hughes et al. (1998) could be identified with VLA 
sources from the 1.4 and 8.5 GHz catalogues of Richards et al. (1998) if the 
true positional errors in SCUBA source positions were significantly in excess 
of the nominal $\theta_{\rm FWHM}/ 2 (S/N)$ values. On the basis of 
associations between VLA J123656+621207 and HDF850.1, and between VLA 
J123649+621313 and HDF850.4, he advocated
that the native SCUBA coordinate frame of Hughes et al. (1998) be shifted 4.8 
arcsec to the West and 3.8 arcsec to the South. However,
applying this offset has the effect of moving the two SCUBA sources
that were initially the closest to their putative VLA identification much
further from them: HDF850.2 and HDF850.6 are 4.2 and 2.9 arcsec, respectively,
away from their nearest VLA source in the native SCUBA reference frame, but
are both $\sim$9 arcsec away once the shift advocated by Richards (1999) is
applied, because their positions in Table 1 are already to the South and West 
of their supposed radio counterparts.
In fact, no translation or rotation of coordinate frames can produce a good 
match between the positions of HDF850.1, HDF850.2, HDF850.4, HDF850.6 and 
HDF850.7 and all five of their respective possible radio counterparts, 
suggesting that there is not a systematic error in the native SCUBA coordinate
frame of the sort deduced by Richards (1999) on the basis of the proximity
of radio and submillimetre sources in the HDF.

\begin{figure}
\epsfig {file=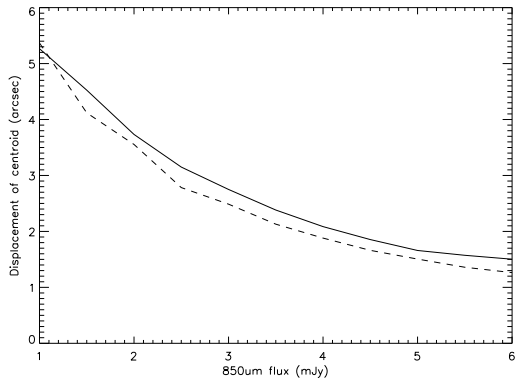,width=7.5cm,angle=0}
\caption{The effect of confusion noise on the positions of sources
extracted from our 850$\mu$m map. The solid (dashed) line shows the
mean (median) value of the distance between input and output source 
positions, as a function of the input flux of the simulated source.}
\label{fig:absdev}

\end{figure}

Another possibility is that the confusion noise in the sub-mm map has
significantly affected the positions of the sources we extracted from it,
so we tested this using simulations. We selected six locations
in the map well away from the locations of our eight detected sources, and
defined at each a 5$\times$5 square grid of positions, with 6 arcsec 
separation between adjacent rows/columns. Then, one
at a time, we placed a simulated source into the map at a grid position, and
ran our source extraction routine on the resulting map, noting the 
displacement between the 
extracted source position and its input location. This process was repeated
for simulated sources of a range of 850$\mu$m fluxes, so that we could measure
how the effect of confusion noise varied with the S/N of the source detection.
The results of this procedure are summarised by Figure~\ref{fig:absdev} which
plots the mean and median values of the distance between input and output 
source positions, as a function of the input flux of the simulated source.
Assuming that there are no systematic differences between the input
and output fluxes, 
we
may then use our simulation results to estimate, for each of the five sources
with putative radio counterparts, the probability that confusion noise could
produce a positional offset as large as that required to reconcile the 
distance between the SCUBA and VLA positions. We find that, for HDF850.1, 
HDF850.2, HDF850.4, HDF850.6 and HDF850.7, the SCUBA--VLA offset corresponds 
to the 97th, 86th, 44th, 91st and 95th percentile, respectively, of the 
distribution of displacements produced by confusion at the relevant flux
level. 

It is perhaps conceivable that by avoiding bright sources in the map,
our simulated sources will have suppressed confusion noise (if the
clustering of the point sources is sufficiently strong). Indeed, the
simulations of Eales et al. (2000) find large astrometric shifts,
albeit with very different mapping and source extraction
algorithms. It is worth noting that the pointing solution of the JCMT
has occasionally required revising with the appearance and discovery
of new systematics (e.g. pointing shifts during transit), and although
well-modelled the causes of these shifts are not all well-understood. 
Nevertheless, in the current 
absence of evidence to the contrary, we will
treat our simulations as accurately reflecting the astrometric
uncertainties in our discussion of identifications. 

We shall discuss below how the low $P_0$ values for the five possible 
VLA--SCUBA associations might have arisen, if these associations are incorrect,
but we conclude from this subsection that neither a systematic shift in 
astrometric
reference frame, nor the effect of confusion noise in our deep 850$\mu$m
map, is sufficient to reconcile the displacements between the SCUBA and
VLA positions of all of this set of five submillimetre sources and  their putative
radio counterparts. Furthermore, the accurate position for HDF850.1 measured
by the IRAM interferometer (Downes et al. 1999) provides one case where we
know for certain that the submillimetre source is not coincident with the
nearby VLA source.

\subsection{Methods for quantifying the reliability of
identifications}\label{sec:associations_reliability}

The reliability of associations made between sources in one catalogue and 
objects in another are typically quantified using a simple Poisson method, 
working solely on proximity (as in 
computing the $P_0$ values above), or  the likelihood ratio method, 
(e.g. Sutherland \& Saunders 1992) which allows the inclusion of
knowledge about the properties of the population under study
(e.g. flux or colour distributions, derived either from prior work or from the
data themselves, if the sample size is large enough) enabling, for example,
the further of two possible objects to be favoured if it matches more
closely the properties of the expected counterpart being sought.

When the surface density of the object catalogue and/or the search
radius is large, the simple 
expression for $P_0$ quoted above does not give a good estimate of the
probability of such an extreme event (i.e. that an object so bright should
be found so close to the source) occurring by chance in the Poisson model,
as shown by Downes et al. (1986). If one searches a catalogue with magnitude
limit $m_{\rm lim}$ out to a radius $r_{\rm S}$ from a  source position,
then the expected number of events as extreme as an association with an
object of magnitude $m$ at a distance $d$ from the source is not simply
equal to $E_0=\pi d^2 N( \leq m)$, where $N(\leq m)$ is the surface density 
of objects at least as bright as magnitude $m$, and so the Poisson model 
probability of this event occurring
by chance is not $P_0=1-\exp(-E_0)$. This is because {\em a posteriori\/}
probabilities at least as low as $P_0$ could have also been produced by 
there being objects brighter than $m$ lying at distances greater than $d$ 
from the source, or by objects fainter than $m$ lying closer than $d$. The
true {\em a priori\/} expected number of events as extreme as the putative
association is then $E=E_0+E_1+E_2$, where $E_1$ and $E_2$ account for
brighter/farther and fainter/closer objects respectively: {\em i.e.\/}
\mbox{$E_1=\int_d^{r_{\rm S}} N[\leq m_{\rm max}(r)] 2\pi r dr$} and 
\mbox{$E_2=\int_m^{m_{\rm lim}} n(m) dm \int_0^{r_{\rm max}(m)} 2 \pi r dr$}, 
where $n(m)$ is the differential number density and where
\mbox{$N[ \leq m_{\rm max}(r)]$} and $r_{\rm max}(m)$ are defined such that 
$\pi r^2$ \mbox{$N[ \leq m_{\rm max}(r)]$}$=E_0$ and 
$\pi N(\leq m) r^2_{\rm max}(m)=E_0$, respectively.

As shown by Downes et al. (1986), this means that the {\em a priori\/}
probability is given by $P=1-\exp(-E)$, where $E$ takes the form
$E=E_0[1+ \ln (E_{\rm c}/E_0)$, with $E_{\rm c}=\pi r^2_{\rm S} N(\leq 
M_{\rm lim})$. This reduces to $P=P_0$ in the limit $E_{\rm c} << E_0$,
which holds for the case of the Richards (2000) radio survey 
(so we were correct in quoting just $P_0$ values above), but does not
hold for deep optical surveys of the HDF/HFF, given their much
higher  $N(\leq M_{\rm lim})$ values: it is clear that, for such surveys, the
exact $P$ values obtained will depend on the value of $r_{\rm S}$ used,
and that a sensible choice for that must be made, on the basis of the
assumed astrometric accuracy of the source catalogue.

A similar choice must be made when using the likelihood ratio method 
(e.g. Sutherland \& Saunders 1992), as that explicitly includes a model
for the probability distribution of offsets between source and object
positions, which typically reduces to the positional error distribution
of the source population. The likelihood ratio is defined by 
\mbox{$LR(f,x,y) = q(f) \cdot e(x,y)/ n(f)$},
where $e(x,y)$ is the probability distribution for positional offsets,
$(x,y)$, between source and object [normalized so that $\int e(x,y)
\; {\rm d}x \; {\rm d}y =1$ with the integral being taken over all
space], $n(f)$ is the surface density of objects per unit interval in
flux, $f$, and $q(f)$ is the probability distribution function for 
an ensemble of sources, measured in the same passband in which the
object catalogue is defined. More generally, $q(f)$ and $n(f)$ can
be replaced by the corresponding quantities for some other property
of the source and object populations, such as their colours, photometric
redshifts, etc.  	 

The quantity $q(f)$ is unknown unless associations have previously been
found between an ensemble of similar sources and a catalogue of
similar objects, or unless the samples under study are sufficiently large
that $q(f)$ can be estimated with sufficient accuracy from the data themselves.
Neither of these conditions are met when, as here, one is seeking associations
for a restricted number of sources drawn from a  population about which 
little is known. In similar circumstances, seeking associations for {\em ISO\/}
sources in the HDF and HDF-S, Mann et al. (1997, 2002) took a constant
value for $q(f)$, after comparing the magnitude distributions of objects near 
source positions to that for the object catalogue as a whole, thereby deducing
a peak in $q(f)$ at $I \sim 21$. This told them that the effect of their 
taking a constant $q(f)$ would be to underestimate the likelihood ratio for 
associations made with objects near that magnitude, which they could bear in 
mind should there be such an object vying with another of a different magnitude
for selection as the likeliest counterpart of a given source; in both studies,
Mann et al. (1997, 2002) found no such cases, justifying their use of a 
constant $q(f)$.

In this case, the number of SCUBA sources is so low that nothing about $q(f)$
can be deduced from comparing the magnitude distributions in this way, but,
since we cannot estimate it well from previous work, either, we are forced to 
take it to be a constant if we are to use the likelihood ratio method at all. 
This leaves the $LR$ value undetermined up to a constant factor, so
we must also follow Mann et al. (1997, 2002) in quantifying the reliability of
associations made via this modified likelihood ratio method through the use
of simulations, rather than using the algebraic methods of Sutherland \& 
Saunders (1992) and/or Rutledge et al. (2000), which can only be applied when 
correctly normalised $LR$ values are available. In this case we compute, for
each association, the probability, $P_{\rm ran}$,
that a fictitious source placed at a random location in the region covered by
the object catalogue would yield an $LR$ value through association with any
object in that catalogue as high as that for the putative identification. 

With $q(f)$ taken to be a constant, the modified likelihood ratio method,
coupled with simulations to compute $P_{\rm ran}$, employs no additional
information than the Poisson method, so the $P$ and $P_{\rm ran}$ values
they yield for a particular association should be similar, given consistent
choices for the search radius $r_{\rm S}$ in the Poisson model and the value
of $\sigma$ in the Gaussian $e(x,y)$ positional error model used in the
likelihood ratio method. One slight difference is that our likelihood ratio
method uses $n(f)$, the differential flux distribution of the object
catalogue, while the Poisson model employs the integral distribution,
$N(\leq m)$. The use of the former may yield more robust results for bright 
optical IDs in situations where the image analyser producing the object 
catalogue splits bright objects into multiple ``children'' in a manner that
is not readily accounted for across a sizeable catalogue, like that of 
Williams et al. (1996). To help assess the effect this might have on
our results, we consider the SExtractor (Bertin \& Arnouts 1996) 
catalogue available from the HDF--S WWW page 
({\tt www.stsci.edu/ftp/science/hdfsouth/catalogs.html})
and which we denote by N98, in addition the original
catalogue of Williams et al. (1996, denoted W96), which was produced by a 
modified version of FOCAS (Jarvis \& Tyson 1981)
 and the much shallower catalogue
of Barger et al. (1999c, B99). In general, we do find that the likelihood
ratio model does produce more consistent results from the different 
catalogues, that the Poisson method yields higher random probabilities
than the likelihood ratio method, and that the differences between results
are greater at brighter magnitudes, all consistent with the idea that these
effects arise from the imperfectly--corrected over--splitting of bright
objects.

\subsection{Association results for individual
sources}\label{sec:associations_individual}

In this subsection, we quote both $P$ and 
$P_{\rm ran}$ values for possible identifications of individual SCUBA 
sources with objects in optical and radio catalogues. When interpreting these 
values, the reader should bear in mind the assumptions behind them, both
those inherent to the two methods [e.g. that the object population is 
unclustered] and those specific to their implementation here (e.g. taking
$q(f)$ to be a constant, the choices made for $r_{\rm S}$ and $\sigma$,
based on the estimated astrometric accuracy of the source positions, and
possible object--splitting problems at bright magnitudes), and should note
that the results for particular values of $r_{\rm S}$ and $\sigma$ are
not always directly comparable.

\subsubsection{HDF850.1}

For the present purposes, we assume that the correct position of this 
source is that of the IRAM 1.3mm source detected by Downes et al. (1999).
That is a distance of 1.9 arcsec from the position quoted in Table 1,
corresponding to the 69th percentile of the distribution of positional
offsets estimated in the simulations of
Section~\ref{sec:associations_scuba_vla}. 
We refer the reader to Downes et al. (1999) for a more detailed discussion 
of the possible associations with this source; nothing in our analysis
conflicts with their conclusions, although we do present here information
from studies of the HDF region that have appeared since the publication
of that paper.

The sub--arcsecond accuracy of the IRAM position confirms that the submm source
is not the same object as the radio/X--ray source close to optical galaxy
3-659.1 (at a distance of 4.7 arcsec, that is at the 95th percentile of
the distribution of positional offsets), although its position is consistent 
(Downes et al. 1999) with 
the tentative (4.5$\sigma$ at 3cm) VLA source 3651+1226 
in the supplementary list of Richards et al. (1998). At first sight there 
are two plausible
optical counterparts: 3-586.0 is an $I_{\rm 814}(AB) \simeq 24$
galaxy 1.0 arcsec away (34th percentile), while 3-593.1 is slightly closer
(0.8 arcsec, 27th percentile) but fainter $I_{\rm 814}(AB) \simeq 26$.
For 3-593.1, the likelihood ratio method yields $P_{\rm ran}$ values of 0.15 
and 0.14 using  the W96 and N98 catalogues, respectively, for an assumed 
SCUBA positional error distribution with $\sigma$=1 arcsec, rising to 0.36
and 0.39 for $\sigma$=2 arcsec, while for 3-586.0 the corresponding 
$P_{\rm ran}$ values are 0.09, 0.14, 0.05 ($\sigma$=1 arcsec) for the
W96, N98 and B99 catalogues, and 0.20, 0.39, 0.13 for $\sigma$=2 arcsec: the
Poisson model yields $P$=0.2--0.3 for both galaxies with both the W96 and
N98 data, with $r_{\rm S}=3$ arcsec.

Downes et al. (1999) note that, with the appearance of an elliptical galaxy
and with photometric redshift estimates in the range \mbox{$1.0 \leq z \leq
1.2$}, 3-586.0 is highly unlikely to be the source of the submm/mm
dust emission detected by SCUBA and IRAM. Photometric redshift estimates
for 3-593.1 are around $z \sim 1.7$ (matching an Scd template), and 
Downes et al. (1999) show that,
taking that value, the SED of HDF850.1/3-593.1 would be similar to the
ultraluminous infrared galaxy VII Zw 31, although it could be a lower
luminosity source gravitationally lensed by 3-586.0 (which would be 
consistent with the lower $P_{\rm ran}$ values for the association of
HDF850.1 with that galaxy), while the lack of a significant $U_{\rm 300}$
detection means that it could lie at a higher redshift: Downes et al. 
(1999) note that the photometric redshift methods of Fernandez-Soto
and Rowan--Robinson both yield (less well favoured, but not wholly
implausible) local probability maxima in the range \mbox{$2.5 \leq z \leq 3$}, 
although the limit on its $U_{\rm 300}$ magnitude is not sufficiently
tight to be sure that it would satisfy the colour selection criteria of
Madau et al. (1996) for \mbox{$2 \leq z \leq 3.5$} galaxies. 

Further redshifts constraints may be deduced from $\alpha^{350}_{1.4}$, the
350 GHz to 1.4 GHz spectral index defined by Carilli \& Yun (1999, 2000), 
under the assumption that both the submillimetre and decimetric radio
luminosities of these sources are proportional to the rate at which they
form massive stars (e.g. Condon 1992). If we assume that HDF850.1 is 
the counterpart of the 8.5 GHz source VLA 3651+1226 (Richards et al. 1998),
and we assume a radio spectral index of $\alpha = -0.8$, then we deduce
a 1.4 GHz flux of 28$\mu$Jy. From the ratio of this and its 850$\mu$m
flux of 5.6mJy, we calculate that $\alpha^{350}_{1.4}=0.96$ which, from
Carilli \& Yun (2000) implies $2.25 \la z \la 4$, while, if we use the
analogous 
relationship derived by Barger, Cowie \& Richards (2000) from a fit to
the SED of Arp220, we obtain $z=2.89$. The flux limit of the
Richards (2000) 1.4 GHz catalogue is 40$\mu$Jy, so, if we disregard the
tentative association of HDF850.1 with VLA 3651+1226, we would deduce
$\alpha^{350}_{1.4} \geq 0.89$, implying $z \ga 1.75$ (Carilli \& Yun 
2000), not too inconsistent with the photometric redshift estimates for
3-593.1 discussed above.

Further evidence supporting the identification HDF850.1 with 
VLA 3651+1226
comes with the convincing detection of a very red ($I-K>5.2$) host
galaxy coincident with the IRAM and VLA positions (Dunlop et
al. 2002). The proximity of 
the elliptical galaxy 3-586.0 implies a gravitational lens
amplification of up to $6.4$, discounting pathological lens/source 
alignments. 
On the basis of fits to all the existing multi-wavelength data, Dunlop
et al. (2002) quote a redshift estimate of $z=4.1\pm0.5$, consistent
with the radio/sub-mm limits quoted above. 

Although 3-586.0 is already ruled out as the source of the
sub-mm emission, it remains associated in the sense of providing
gravitational lens amplification of HDF850.1. 
This is a salutory lesson for the problem of associations with sub-mm
point sources: low values of the $P$ and $P_{\rm ran}$ statitstics may
indicate any of a number of physical associations with the sub-mm
source, such as gravitational lensing, large-scale structure, or 
the physical co-location of the multiwavelength emission. 


\subsubsection{HDF850.2}

This source lies outside the HDF, so we can only seek optical associations
with objects in the much shallower catalogue of Barger et al. (1999c), with
limiting magnitude $I \sim 24$, and to this depth we find no plausible optical
ID for HDF850.2: the nearest is the $I \sim 23$ galaxy 36564 
1209, which lies 5.6 arcsec away from the SCUBA position, yielding 
$P_{\rm ran}$ values in the range 0.6--0.7 for a positional accuracy of
2--3 arcsec, as estimated from the simulations of
Section~\ref{sec:associations_scuba_vla}. 
The radio source VLA J123656+621207 lies 4.2 arcsec away from the SCUBA
position, corresponding to the 90th percentile of the distribution of 
offsets in the simulations of Section~\ref{sec:associations_scuba_vla}. 
Barger et al. (2000) report an 850$\mu$m flux of 2.5$\pm$0.7 mJy at that
position, consistent with the value given in our Table 1, though this
is also consistent with the SCUBA source being separated from the
radio source by 5.6''. 
Barger et al. (2000) list the detection of an optical counterpart
for the radio source with B=26.2, in addition to upper limits in a series of
other bands (HK$^{\prime} > 22.6$, I$> 25.3$, R$> 26.6$, V$> 26.4$, 
U$^{\prime} > 25.8$), while, on the basis of their radio/submillimetre 
spectral index method they deduce a redshift of $z=1.8^{+0.7}_{-0.5}$.

\subsubsection{HDF850.3}

This SCUBA source has one of the larger positional uncertainties amongst
our sample because of the low S/N of its detection: indeed, as noted above,
it does not satisfy our revised source extraction criteria, and we include
it here only for consistency with Hughes et al. (1998). The position we
deduce for this source is about 5 arcsec to the north--east of that 
determined by Hughes et al. (1998), and this shift makes their preferred
ID (1-34.2) much less likely: we compute $P$ and $P_{\rm ran}$ values 
in excess of 0.9, indicating that it is highly likely that an $I_{\rm 814}(AB)
\sim 24$ galaxy should be found within 5 arcsec of a SCUBA position by chance.
Much lower random probabilities are computed for the brighter [$I_{\rm 814}(AB)
\sim 21$] galaxy 1-34.1, which is 3.4 arcsec from the SCUBA position from
Table 1, corresponding to the 24th percentile in the simulated offset
distribution: for the W96, N98 and B99 catalogues, we compute $P_{\rm ran}$=
0.16, 0.20 and 0.10, respectively, for $\sigma$=2 arcsec, and 0.16, 0.29 and
0.14, for $\sigma$=5 arcsec. Richards (1999) reports that this galaxy has
a 3$\sigma$ radio detection at 1.4GHz, and that its level of 23$\mu$Jy
is more than an order of magnitude lower than what would be expected on the
basis of the FIR--radio correlation, should the SCUBA emission be associated
with 1-34.1, given its spectroscopic redshift of 0.485 (Phillips et al. 1997).
In fact, it would need to be a factor of $\sim$30 times higher to match the
canonical $\alpha^{350}_{1.4}$ relation of Carilli \& Yun (1999) for a
$z=0.485$ galaxy, while, conversely, the Carilli \& Yun (2000) relation
suggests that the true association lies at $0.9 \la z \la 2.25$.

We conclude that the only plausible optical association for this source,
should the SCUBA source be real, is 1-34.1, but that we do not judge this to 
be a very
reliable identification, both because there is a moderately high probability 
(0.1--0.2) of the proximity of such a source occurring at random, and also
because the low redshift of this galaxy should lead to a 1.4GHz
flux $\sim$30 times higher than is detected (cf. a factor of $\sim$2 scatter
in the $\alpha^{350}_{1.4}$ relation): the $\alpha^{350}_{1.4}$ relation
implies  $0.9 \la z \la 2.25$ for this source.  

\subsubsection{HDF850.4}

Hughes et al. (1998) associated this source with the $I_{\rm 814}(AB)=23$
galaxy 2-339.0, which lies only 0.8 arcsec away from the SCUBA source position
listed in Table 1. This corresponds to the 2nd percentile of the simulated
offset distribution, and both the likelihood ratio and Poisson methods yield
probabilities of less than 0.1 that this association should occur by chance,
given a positional accuracy of 2--3 arcsec. Richards (1999) prefers an 
association of this source with VLA J123649+621313, which lies 5.3 arcsec
away (50th percentile) and is clearly associated with the galaxy 2-264.1,
which is also the optical counterpart of an X--ray source from Brandt et al.
(2001).
Barger, Cowie \& Richards (2000) measure an 850$\mu$m flux of 1.0$\pm$0.6
mJy at the position of VLA J123649+621313, which does not itself constitute
a significant detection, but is not inconsistent with the flux quoted for
HDF850.4 in Table 1 above.

The SEDs resulting from these two associations have been discussed by 
Cooray (1999), who compares them with an Arp220 model shifted, in the case
of 2-264.1, to $z=0.475$ (the spectroscopic redshift measured for it by
Cohen et al. 1996), and to $z=0.9$ for 2-339.0, for which Hughes et al. (1998)
report photometric redshifts in the range 0.74-0.88. Cooray (1999) finds that
the Arp220 model gives a good fit to the optical--to--submm SED data resulting
from both possible associations with HDF850.4, but that the identification
with 2-264.1 has the additional support as this agreement is extended into
the radio through the detection of  VLA J123649+621313 at a flux consistent
with the Arp220 model. Conversely, and as noted by Richards (1999), the lack of
a 1.4GHz detection of 2-399.0 is difficult to square with the radio--FIR
correlation: the $\alpha^{350}_{1.4}$ relation of Carilli \& Yun (1999)
would predict a 1.4 GHz flux of $\sim$280$\mu$Jy for a galaxy at $z=0.9$,
seven times the flux limit of the Richards (2000) catalogue.
 As noted by Brandt et al. (2001), the detection of 2-264.1 
by Chandra is consistent with its being a powerful starburst (its soft
X--ray luminosity is $\sim$5 times that of M82: Griffiths et al. 2000) rather
than an AGN,  while its 1.4--to-8.5GHz spectral index ($\alpha=0.72 \pm0.15$)
is inconclusive.

In summary,  the optical data strongly favour 2-399.0 over 2-264.1, 
but the lack of a radio detection for 2-399.0 is very difficult to reconcile 
with an Arp220--like SED at the redshift assumed by Hughes et al. (1998). One possible 
reason why there is a lower probability for the random association with
2-264.1 may be that the proximity of HDF850.4 and HDF850.5 has shifted the centroid of
the former to the East, which takes it further from 2-264.1, thereby making
that association seem less secure than it should. On balance, we favour the
associated with 2-264.1, but cannot rule out that with 2-399.0 or 
the possibility that HDF850.4 has no optical counterpart to the depth of
the W96 catalogue.

\subsubsection{HDF850.5}

As with HDF850.4, the proximity of these two sources might have
significantly moved the centroid of this SCUBA source, however, in this 
case, the shift would have been towards the most likely optical ID,
2-404.0, so, if anything, we are likely to have over--estimated the
plausibility of its being the optical counterpart to HDF850.5. Even with that
caveat, 2-404.0 does not appear a very secure identification: it lies
6.6 arcsec from the position of HDF850.5 given in Table 1, which corresponds
to the 67th percentile in the offset distribution, for a source of its
flux, and its redshift ($z=0.199$ Lanzetta et al. 1996) is inconsistent
with the lower limit of $z \ga 0.75$ deduced by the method of Carilli
\& Yun (2000) on the basis of the lack of a 1.4 GHz detection above the 
40$\mu$Jy limit of the Richards (2000) catalogue.
Hughes et al. (1998) favoured an association with the  
$I_{\rm 814}(AB)=29$ galaxy 2-426.0, which is only 0.9 arcsec from the
SCUBA position, however, with a likely positional accuracy of no better
than $\sim$3 arcsec, there is a high probability ($\geq 0.7$) of that
occuring by chance, and that would only be increased were the true position
of HDF850.5 to lie further to the East than that listed in Table 1, should
the proximity of HDF850.4 have caused a significant shift in the position
of both sources. We conclude that there is no reliable identification for
this source.

\subsubsection{HDF850.6}

This is the first of the new sources, detected beyond the region
studied by Hughes et al. (1998). It lies outside the HDF, so our
optical information is limited to the shallow catalogue of Barger
et al. (1999c). The nearest object in that catalogue is 37018 1143,
an $I=21.87$ galaxy at a distance of 4.4 arcsec: that yields 
likelihood ratio random probabilities of $P_{\rm ran}=0.3-0.4$
for $\sigma=2-5$ arcsec, and Poisson probabilities in excess of
0.2 for $r_s \geq 5$ arcsec, so this is not a likely identification.

HDF850.6 lies 2.9 arcsec from the radio source
VLA J123701+621146: this distance corresponds to the 93rd percentile
of the offset distribution for a source of this flux, but this region
of our SCUBA map is unusually noisy, so the 850$\mu$m detection is
made at S/N $<$4, and it is quite likely that we have underestimated
its true positional uncertainty. We adopt the radio source as the most 
likely identification. Barger et al. (2000) measure an
850$\mu$m flux of 4.7$\pm$2.1 mJy at the location of VLA J123701+621146,
which is consistent with that quoted in our Table 1 above, though as
with HDF850.2 this does not itself prove the radio and sub-mm sources
are cospatial. 
This radio
source is itself associated (Alexander et al. 2001) with an Extremely
Red Object (ERO) with I-K$ > 5$, for which Cohen et al. (2000) report
a spectroscopic redshift of $z=0.884$, on the basis of their identification
of a single detected emission line as $[{\sc oii}] \lambda 3727$; this 
is just consistent with the lower limit to the source redshift determined
by the radio/submillimetre spectral index method of Carilli \& Yun (2000),
on the basis of the association of HDF850.6 with VLA J123701+621146, which
implies $0.9 \la z \la 2.25$. 
Alexander et al. (2001) also detect X--ray flux for this 
source in the Chandra map of Brandt et al. (2001) (it is not included in
the catalogue of Brandt et al. 2001, because it falls below their
significance threshold; Alexander et al. 2001 accept sources with a lower
significance, where they are coincident with EROs) and they
show that the extant data for this galaxy (they
plot the 850$\mu$m flux of Barger et al. 2000) are reasonably well fit
by shifting the Silva et al. (1998) SED model for Arp220 to $z=0.884$ and
dividing its luminosity by 2.2. We conclude that this is the most likely 
association for HDF850.6, but cannot rule out the possibility that this
SCUBA source has no optical counterpart to the limit of the Barger et
al. (1999c) optical catalogue.

\subsubsection{HDF850.7}

For HDF850.7, the nearest candidate HFF identification is a $z=1.219$ 
(Barger et al. 2000), $I=22.3$ galaxy at a distance of 
5 arcsec, which is the 98th percentile of the simulated offset distribution,
given the source flux. This large displacement leads to a high random
probability ($>$0.30 from both Poisson and likelihood ratio methods), 
although, as with HDF850.6, this is a noisy region
of our map, so we are certainly under--estimating the positional uncertainity.
(The radio source is offset by $2.5\sigma$ from the SCUBA centroid,
assuming a sub-mm positional uncertainty of $\theta_{\rm
FWHM}/(2*S/N)$.) 
This galaxy is a Chandra source and is identified with the ISO source 
HDF\_PM3\_3 (Aussel et al. 1999) with a $15\mu$m flux of 
$363\mu$Jy. It is also the radio source VLA J123634+621241, with a $1.4$GHz 
flux of $230\pm13.8\mu$Jy, from the Richards et al. (2000) catalogue. If this 
were the correct identification, the radio/submillimetre flux ratio method of
Carilli \& Yun (2000) would imply a redshift between $z=0.6$ and $z=1.7$, 
consistent with the spectroscopic redshift above. Barger et al. (2000)
measure an 850$\mu$m flux of 2.1$\pm$2.3mJy at the position of 
VLA J123634+621241, which is not inconsistent with the flux for HDF850.7
listed in Table 1. 
If the association is not accepted, and there is no radio counterpart for
HDF850.7 down to the 1.4 GHz flux limit of 40$\mu$Jy for the Richards (2000)
catalogue, then the redshift range implied by the Carilli \& Yun (2000)
method would be $1.75 \la z \la 4$. 
We conclude that there is no secure
identification for this source: an association with the $z=1.219$ galaxy
identified with the radio source VLA J123634+621241 cannot be ruled out
on the basis of the 850$\mu$m flux measured at that position by Barger
et al. (2000), although it has a high probability of being a chance
occurence (albeit under a simplistic noise model), and it is possible that
HDF850.7 lies at $z \ga 1.75$, with no optical counterpart to the depth of 
the W96 catalogue.

\subsubsection{HDF850.8}

This SCUBA source is 2.0 arcsec (13th percentile) away from 
2-736.1, which seems to be one an interacting pair of galaxies, which 
has a spectroscopic redshift of $z=1.355$ (Cohen et al. 1996). The
random probabilities from the likelihood ratio method are quite low
0.05, 0.14, 0.07 for W96, N98 and B99 for $\sigma=2$ arcsec, and the
lack of a radio detection is consistent with this redshift, given the
radio/submm correlation method of Carilli \& Yun (1999), which 
implies $z \ga 0.9$. We conclude that this is a likely identification
for HDF850.8.

\subsection{Summary of association results}

Despite the appearance of a great deal more multi--wavelength data for the
HDF/HFF region since the publication of our initial analysis of the
SCUBA HDF (Hughes et al. 1998), it remains the case that we do not have secure
optical/near--IR IDs for the majority of the 850$\mu$m sources in this field. 
Our two most secure such IDs (those for HDF850.8 and HDF850.1) lie at
$z \geq 1$, 
one source (HDF850.2) seems securely identified with a VLA source in
an optically-blank 
field believed to lie at $1.3 \la z \la 2.5$, 
while HDF850.4 (and possibly also HDF850.6 and HDF850.7) appear to be
associated with  
$z<1.5$ galaxies detected in both the radio and the X-ray
(although none of the cases is beyond doubt) 
and with SEDs well fit by 
an appropriately redshift Arp 220 SED. The final
two sources (HDF850.3, HDF850.5) have no likely
identification at all,  
only $z \ga 0.75$ redshift constraints, based on non--detections in
the radio. The available redshifts and redshift constraints are listed 
in table \ref{tab:ids}, and are consistent with source count model
predictions (e.g. Rowan-Robinson 2001). 

It is interesting to note that, if we had adopted the strategy of
selecting 
optically--faint ($I \geq 24$) radio sources as likely SCUBA sources,
as advocated by 
Barger et al. (2000) and Chapman et al. (2001a,b), we would only have
recovered 
two out of the eight sources, well below the 70 per cent success rate
claimed for this 
technique by Chapman et al. (2001b). {\it Provided} we have not
underestimated our astrometric uncertainties, this implies that this
pre--selection method may lead 
to incompleteness in detecting the submillimetre source
population. Furthermore, 
giving optically--faint radio sources a high prior probability of
being the counterparts 
of SCUBA sources can lead to mistaken identifications. For example,
the radio source  
VLA J123651.7+621221, which lies 4.7 arcsec from the position for
HDF850.1 given in  
Table 1, has $I > 25.3$ (Barger et al. 2000) and yields a Poisson
probability of $P_0=0.05$ 
of being found so close to the SCUBA position by chance: if we
believed that optically--faint  
radio sources are likely to be associated with SCUBA sources, we might
have accepted this 
as a very plausible identification, but the high positional accuracy
of the IRAM detection 
of this source (Downes et al. 1999) tells us that VLA J123651.7+621221
cannot be the correct  ID. The discovery of a lensed near-infrared
counterpart (Dunlop et al. 2002) at the IRAM position further argues
against this ID. 

If we accept that the five original SCUBA sources from Hughes et al. (1998) 
are not all robustly identified with VLA sources from the Richards (2000)
catalogue -- in fact,  
we argue above that only two are -- how can we 
understand the result of Section~\ref{sec:associations_scuba_vla} that
all five of them yield $P_0 \leq 0.05$ 
for association with objects in that catalogue?  
It is clear that the confusion noise in our deep SCUBA image does have
a significant impact on the accuracy with which we can estimate the
positions of extracted 
sources. However, this cannot be the main factor: even if our SCUBA
source positions 
are offset by a few arcsec by this effect (as indicated by
Fig.~\ref{fig:absdev}), the true 
$P_0$ values must still be very low, because the radio catalogue is so
sparse ($\la$1 source 
per square arcmin). The situation we observe is reminiscent of that
reported by Almaini 
et al. (2001), who noted that SCUBA sources in the ELAIS N2 field
appear to trace the same 
large--scale structure as objects detected in their deep Chandra image
of the same field, 
but not to be identified with them. Almaini et al. (2001) sought to
explain their result 
by postulating that the SCUBA phase and the AGN phase in a galaxy's
evolution are not 
simultaneous, so that within the galaxy population at $z \sim 2$, say,
there will be galaxies 
exhibiting both sets of classes of behaviour.

While qualitatively similar, there is a quantitative 
difference in scale between that result and ours here: the
SCUBA--Chandra pairs discussed 
by Almaini et al. (2001) are typically tens of arcsec apart (their
cross--correlation function 
is significantly positive to a separation of $\sim$100 arcsec), while
our SCUBA--VLA pairs 
are separated by less than 6 arcsec. One solution is that the deeper
data available in the 
HDF/HFF region are probing further down the luminosity function than
is happening in ELAIS 
N2 (this is certainly the case for the submillimetre and hard X--ray
data, since, in both 
cases, favourable k--corrections mean that the flux--redshift relation
is flat for a  
source of a given luminosity at $z>1$), so the number densities of
detected objects drawn 
from these populations are higher, and, therefore, the mean separation
of nearest neighbours 
will be lower. More fundamentally, the identification of the SCUBA
sources with a population 
exhibiting significant clustering invalidates one of the assumptions
in both the Poisson 
probability estimation method and the assignment of $P_{\rm ran}$
probabilities in our 
implementation of the likelihood ratio association procedure: an
assessment of 
how large an effect this is on the probabilities is very
model--dependent, given existing 
knowledge of the SCUBA source population, so we shall not attempt it here.

\begin{table*}
\begin{tabular}{llllll}
~Name & HDF/HFF ID name& $S_{1.4{~\rm GHz}}$ ($\mu$Jy) & Other Photometry & Candidate ID $z$  \\
      &            &                               &       & (and $z$ constraint from $\alpha^{350}_{1.4}$ )  \\
      &            &                               &                                        &   \\

~HDF850.1   & VLA J123651+621226 & $16\pm4$ &  $I=23.40\pm0.05$ & $z_{\rm phot} \simeq 4.1 \pm 0.5$ \\
 &  & & $H=20.40\pm0.05$ &  ($z\ga 1.7$) \\ 
 &  & & $K=19.39\pm0.03$ & & \\
 &  & & $S_{1.3{\rm mm}}=2.2\pm0.3$mJy & & \\
 &  & & $S_{8.4{\rm GHz}}=7.5\pm2.2\mu$Jy & & \\


 \\
~HDF850.2   & VLA J123656+621207 & $46.2\pm7.9$ & $U^{\prime} > 25.8 $ & \\
 &  & & $B=26.22$ & ($z\ga1.3$)\\  
 &  & & $V > 26.4$ &  \\
 &  & & $R > 26.6$ &  \\
 &  & & $I > 25.3$ &  \\
 &  & & $HK^{\prime} > 22.6$ &  \\
\\

(HDF850.3)  & 1-34.1 (?) & $<40$ &      $U_{300}= 25.24$ & $z_{\rm spec}=0.485$ (?) \\
 &  & & $B_{450}=23.60$  & ($z\ga0.9$) \\ 
 &  & & $V_{606}=22.20$ & \\
 &  & & $I_{814}=21.22$ &  \\
 \\
~HDF850.4   & 2-264.1 & $49.2\pm7.9$ &  $U_{300}=24.92$ &             $z_{\rm spec}=0.475$  \\
 &  & & $B_{450}=23.48$ &  ($z\ga0.5$)\\ 
 &  & & $V_{606}=22.24$ &  \\
 &  & & $I_{814}=21.39$  &  \\
 &  & & $S_{0.5-8keV}=1.5\times10^{-16}$ ergs/s & \\
 &  & & $S_{0.5-2keV}=6.49\times10^{-17}$ ergs/s & \\
 &  & & $S_{2-8keV}=1.71\times10^{-16}$ ergs/s & \\
 \\
~HDF850.5   & -- & $<40$ & & \\
& & & & ($z \ga 0.75$) \\
 \\
~HDF850.6  & VLA J$123701+621146$(?) & $128\pm9.9$ & $U^{\prime} > 25.8$ &  $z_{\rm spec}=0.884$\\
 &  & & $B > 26.6 $ &  ($z\ga0.9$)\\ 
 &  & & $V > 26.4 $ &   \\
 &  & & $R= 25.73 $ &   \\
 &  & & $I=24.81 $ &   \\
 &  & & $HK^{\prime}=20.14 $ &   \\     
 &  & & $S_{15{\mu\rm m}}=15\pm9\mu$Jy &  \\
 &  & & $S_{0.5-8\rm keV}=1.4\times10^{-16}$ ergs/s & \\
 &  & & $S_{0.5-2\rm keV}=4\times10^{-17}$ ergs/s & \\
 \\
~HDF850.7   & VLA J$1236345+621241$(?) & $230\pm13.8$ & $U^{\prime}=24.72$ & $z_{\rm spec}= 1.219$ \\ 
 &  & & $B=24.08$ & ($z\ga0.6$)\\ 
 &  & & $V=24.22$  &  \\
 &  & & $R=23.25$  &  \\
 &  & & $I=22.29$ & \\
 &  & & $HK^{\prime}=19.10$ & \\
 &  & & $S_{15{\mu\rm m}}=363^{+79}_{-38}\mu$Jy &  \\
 \\ 
~HDF850.8   & 2-736.1 & $<40$ &   $U_{300}=23.01$ & $z_{\rm spec}=1.355$ \\
 &  & & $B_{450}=22.38$ & ($z \ga 0.9$) \\
 &  & & $V_{606}=22.27$ &  \\
 &  & & $I_{814}=22.00$  & \\
\end{tabular}
\caption{\label{tab:ids}Multi-wavelength identifications of the sub-mm 
sources in the Hubble Deep Field. A question mark in the  HDF/HFF ID name column denotes
an uncertain identification: see text for details. HST magnitudes
i.e. $(U_{300},B_{450}, V_{606}, I_{814})$ are from Williams et al. (1996)
while the remaining optical/near-infrared magnitudes are from Barger
et al. (2000) and Dunlop et al. (2002). 
Mid-infrared fluxes are from Aussel et al. (1999). 
X-ray fluxes are from Brandt et al. (2001) and Alexander et al. (2001). 
Radio fluxes are from
the catalogue of Richards (2000), redshifts for candidate IDs are denoted by $z_{\rm spec}$
or $z_{\rm phot}$ for spectroscopic measurements or photometric
estimates, respectively, while the redshift constraints in parentheses
are derived from the sources' submillimetre--to--radio spectral index, according 
to the method of Carilli \& Yun (1999, 2000). 
The lower bounds using this relation are more robust than the upper
bounds due to the steeper
slope of the spectral index vs. redshift relation at lower redshifts. 
Also, due to the uncertainty of several radio-sub-mm
cross-identifications (exemplified by the 
cautionary case of HDF850.1, e.g. Dunlop et al. 2002), we only quote lower
limits on the redshift from the submillimetre--to--radio spectral
index. 
}
\end{table*}

\begin{table*}
\begin{tabular}{lllcccc}
Source & \multicolumn{2}{l}{$1.4$ GHz position} & $S_{850}$ ($30''$) & $S_{850}$ ($45''$) & $S_{450}$ & $S_{1.4{\rm GHz}}$\\
 & (J2000)      &     & (mJy)              & (mJy)              & (mJy) & ($\mu$Jy)\\
VLA/WRST & 12 37 01.574 & +62 11 46.62 & $7.3$ $\pm$ $2.0$ & N/A & ~~$-34\pm51$~ & $128\pm9.9$~~\\
VLA/WRST & 12 36 42.098 & +62 13 31.42 & $0.32$ $\pm$ $0.82$ & $2.87\pm0.94$ & ~$-8.8\pm7.7$ & $467\pm24.6$\\
VLA/WRST & 12 36 44.386 & +62 11 33.10 & ~$0.1$ $\pm$ $1.45$ & $3.7\pm1.7$ & ~~~$82\pm61$ & $1290\pm61.2$~~\\

WRST & 12 36 46.284 & +62 12 36.03 & $0.71\pm0.66$ &  $1.67\pm0.72$ & ~~~$2.0\pm4.4$ & $467$ \\
WRST & 12 36 53.050 & +62 11 37.67 & $-0.15\pm0.87$~~ & $2.90\pm1.25$ & ~~~~~$7.1\pm13.1$ & $73$\\

\end{tabular}
\caption{\label{tab:vla} Most significant candidate 
detections of $1.4$ GHz
sources from Richards et al. (2000) and Garrett et al. (2000) 
in the $850\mu$m
maps with the Hughes et al. (1998) sources subtracted. The first
source is identified with HDF850.6 (table \ref{tab:sourcelist}).
The observed
fluxes quoted are 
the flux per beam in the $850\mu$m maps at the positions of the
sources, but are subject to source confusion and blending. The
noise estimates are the instrumental and sky noise only, and do not
include confusion noise or calibration error. 
}
\end{table*}

\begin{table*}
\begin{tabular}{llllllll}
NICMOS & WFPC2  & z & $S^{\rm pred}_{850}$ & SFR$^{\rm pred}$ & $L^{\rm pred}_{\rm bol}$ & $S^{\rm obs}_{850}$ & $S^{\rm obs}_{850}$\\
ID     &  ID    &   & (mJy)     & $M_\odot/$yr & $L_\odot$ & $30''$ chop & $45''$ chop\\
        &          &        &         &        &                     &     & \\
166.000 &  4-307.0  &  1.60 &  1.78   &  534.7  &  $1.21\times10^{12}$ & ~~$0.07\pm0.68$ & ~~$0.90\pm0.76$\\
277.211 &  4-186.0  &  1.84 &  1.51   &  375.2  &  $1.07\times10^{12}$ & ~~$1.57\pm0.66$ & ~~$1.20\pm0.76$\\
\end{tabular}
\caption{\label{tab:nicmos}
The brightest 
NICMOS HDF galaxies from Thompson et al. (2000).
These two are also the galaxies with the largest
estimated star 
formation rates and largest estimated bolometric luminosities by
Thompson et al., listed in columns 5 and 6 respectively. The quoted
observed fluxes are from the $850\mu$m maps at each chop throw with
the point sources from Hughes et al. (1998) subtracted. 
The observed fluxes quoted are
the flux per beam in the $850\mu$m maps at the positions of the
sources, but are subject to source confusion and blending. The
noise estimates are the instrumental and sky noise only, and do not
include confusion noise or calibration error. 
}
\end{table*}

\subsection{Cross-correlations with near-IR, VLA, ISOCAM, Chandra, and 
AGN candidates}\label{sec:xcorr}

\subsubsection{Cross-correlation statistics}

We have already stressed the difficulties in extracting further
discrete point sources in the $850\mu$m map, due to the problems of
confusion and blending. Nevertheless, the map still contains many
further real (blended) peaks and positive flux. 
In Peacock et al. (2000) we showed that after subtraction of the point 
sources, the residuals in the $850\mu$m map correlate well with the
positions of HDF Lyman break galaxies. Do these fluctuations also
correlate with other known populations? 
We consider the $15\mu$m ISO sources of 
Aussel et al. (1999) and Serjeant et al. (2002 in preparation); the
$1.4$ GHz catalogue from Richards et al. (2000) and Garrett et
al. (2000), and the subset with optical blank fields from 
Richards et al. (1999); 
the hard X-ray Chandra sources of
Hornschemeier et al. (2000), Brandt et al. (2001) and Alexander et
al. (2001); the AGN and AGN
candidates of Jarvis 
\& MacAlpine (1998), Conti et al. (1999), and 
Richards et al. (2000); and the deep NICMOS sources of Thompson et al
(2000). 

In order to find the best cross-correlation statistics, we made
numerical simulations of SCUBA-HDF maps and tested the distributions
of 
(a) unweighted means of the fluxes at the source 
positions, as performed in Peacock et al. 2000; 
(b) Kolmogorov-Smirnov (KS) test significances, comparing 
the flux distribution at the source positions from each catalogue 
with that of the whole map; (c) noise-weighted coadds of the fluxes at 
the source positions, using the noise estimates derived in equation
\ref{eqn:convol_err}. 
As controls, we
performed the same cross-correlations on 
random positions. We assumed a source count model consistent with
previous sub-mm survey data; the simulation results were not found to
be sensitive to the choice of model. 

The unweighted means correctly 
reproduced the means of the input source fluxes, 
as expected, even where the number of beams per source was of order 
unity.  We also checked the frequency of false 
negatives (failure to detect an underlying signal) and false 
positives (apparent
detection of a cross-correlation signal which was not input into the
simulation).  
The unweighted means and KS tests both gave the expected level of
false positives (provided the number of beams per source was  
$\stackrel{>}{_\sim}1.5$), but suprisingly 
the noise-weighted control coadds gave
far more false positives than expected. 
This may perhaps be due to neglecting the (unknown) level of source 
confusion as a noise term. In the following Sections we therefore 
avoid using noise-weighted coadds. 
As regards false negatives, 
there is a non-negligible level of false negative detections 
for both KS and unweighted means: for example, $<40\%$ of
simulations gave a $>2\sigma$ detection where all the target sources
had $<0.5$ mJy. 

\subsubsection{Hard X-ray sources and AGN}

Neither the CHANDRA mega-second sample nor the AGN cross
correlations 
yielded significant detections at $850\mu$m or $450\mu$m, using 
KS tests and unweighted means. 
We can use this to derive limits on the
average flux-flux ratios in the CHANDRA population, for which we
obtain the $2\sigma$ limit 
$S_{850\mu\rm m}/(S_{\rm X}) < 6\times10^{14}$ 
mJy / (erg s$^{-1}$ cm$^{-2}$) for the $0.5-8$keV band, or
$7\times10^{14}$ for the $2-8$keV band.
This implies the population dominating the hard X-ray background
contributes less than $2$mJy per square arcminute at $850\mu$m,
i.e. $<15\%$ of the $850\mu$m extragalactic background light. 
This 
in turn excludes models where the bulk of the sub-mm population are
dust-enshrouded (Compton-thin) 
AGN, rather than luminous starbursts, as other authors 
have also noted (e.g. Hornschemeier et al. 2000). 

\subsubsection{Very red objects}

We tested the Very Red Object (VRO, $I-K>4$) catalogues of Alexander et
al. (2001). After excluding the ERO in the neighbourhood of HDF850.6,
no significant 
correlation was found with the VROs, whether Chandra-detected or not. 

\subsubsection{Decimetric radio sources}

We next consider the $1.4$ GHz sources in the Hubble Deep Field
detected by the VLA and MERLIN. 
The entire $1.4$ GHz catalogue of Richards et al. (2000) yielded
a marginal ($\sim2\sigma$) cross-correlation in both the $30''$ and
$45''$ chop maps. In the $30''$ chop throw map
this is due to a single detection, identified with HDF850.6. 
In the $45''$ map however,
this source is outside the $850\mu$m coverage and the signal is due
partly to 
a $2.9\pm0.9$ mJy ($3.0\sigma$) detection of the 
$1.4$ GHz source at  12 36 42.098  +62 13 31.42, with radio flux 
of $467 \pm 24\mu$Jy (Table \ref{tab:vla}). Despite a comparable noise
level at this 
position this source is {\it not} detected in the $30''$ chop throw map. 
If this source is real, then the failure to detect this source in the
$30''$ map may be due to a negative noise spike at this position, or
a chance blending with a PSF hole from a source $30''$ distant. 
{\it If} this sub-mm detection 
is indeed real, the radio:sub-mm flux ratio implies $z<0.75$. The
upper bound would be inconsistent with the lack of a $450\mu$m
detection at around $2.5\sigma$, if adopting the lower bound to the
$450:850\mu$m ratio Hughes et al. 1998. 
Barring these
three VLA sources in Table \ref{tab:vla}, the remaining
cross-correlation is weaker 
($1.9\sigma$ in the $45''$ map, and $1.4\sigma$ in $30''$). The
preliminary  deeper $1.4$ GHz
list of Garrett et al. (2000) yielded a slightly more significant
cross-correlation: $2.5\sigma$ and $1.6\sigma$ at $45''$ and $30''$
respectively. The $45''$ result dropped to $1.8\sigma$ 
on exclusion of the candidate VLA source detections in Table
\ref{tab:vla}, and almost all of this tentative signal is due to two
further candidate identifications. These are also listed in Table
\ref{tab:vla}. 

We can also obtain constraints on the $850\mu$m:$1.4$ GHz flux ratio. 
We obtain a limit on the mean $350-1.4$ GHz
spectral index of between $0.25$ and $0.5$, implying that the $1.4$
GHz $\mu$Jy population typically lies at redshifts $z<1$ (Carilli \&
Yun 2000) in agreement with existing spectroscopy. This limit 
is not affected by the exclusion of the candidate detections of $1.4$
GHz sources, although 
the significance of the cross
correlation drops below $2\sigma$ when making this exclusion.

We also attempted a cross-correlation with the subset of $1.4$ GHz
sources which are optical blank fields, from Richards et al. 1999. 
For this subset a $\sim2\sigma$ cross-correlation signal was found, which 
we found to be due entirely to HDF850.6. 

\subsubsection{Mid-infrared sources}

For the ISO sources, a $\sim2\sigma$ cross-correlation
signal was found, and is due to a 
correlation between the ISO sources and structures on
the western edge  
of the $45''$ chop throw $850\mu$m map. Curiously, such a correlation
does not exist in the $30''$ map, despite comparable noise
levels. Such a disagreement between the chop throw maps is not
unexpected in our simulations, as the probability of false negatives
is not negligible (see above). Curiously also, 
this mirrors the lack of $1.4$ GHz
statistical detections in this area of the $30''$ map (see above) as
compared against the $45''$ map. 

\begin{figure*}
  \ForceWidth{5.0in}
   \BoxedEPSF{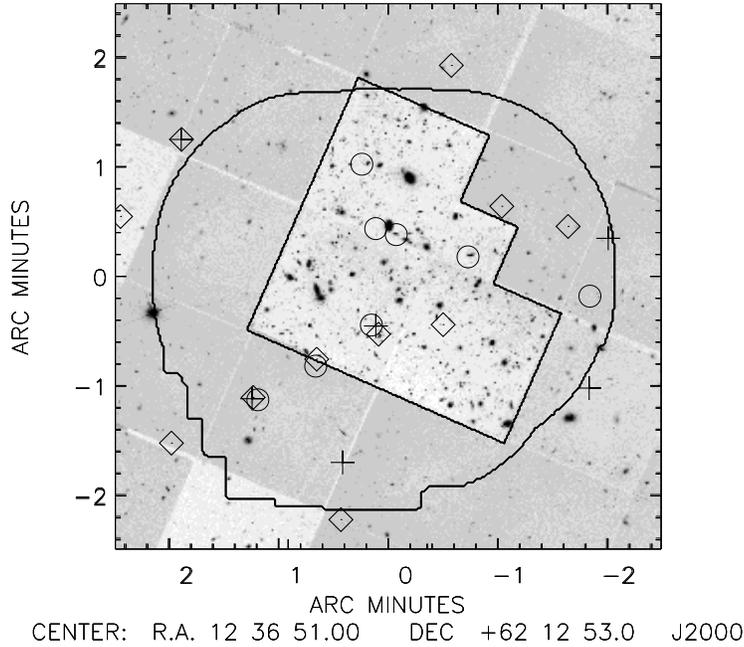}
\caption{\label{fig:scubahdf_sources}
SCUBA-HDF sources (circles) overlaid on a mosaic of Hubble
Flanking Field F814W images. The scale is the same as that of figure
\ref{fig:scubahdf}. The outer contour shows the areal coverage of the
SCUBAHDF $850\mu$m map, after convolution with the beam. The inner
contour shows the areal extent of the Hubble Deep Field North. 
Also plotted are VLA sources with optical 
blank fields (Richards et al. 1999; diamonds),
and Chandra-detected Very Red Objects (Alexander et al. 2001;
crosses; n.b. distinct from the catalogue of Brandt et al. 2001). 
Despite appearances to the contrary, in at least one case 
(HDF850.1) the Chandra and/or VLA sources are demonstrably 
{\it not} identified with the sub-mm source. 
}
\end{figure*}

\subsubsection{K-band sources}

The $282$ NICMOS sources of Thompson et al. (2000) cover an area less
than one arcminute square, in which there are less than $70$ beams at
$450\mu$m and $<20$ at $850\mu$m. This high source density per beam
makes the statistics used above unsuitable. 
Instead, we use the predicted $850\mu$m
fluxes of Thompson et al. to make simulated confused maps for each
chop throw, and use an 
M82 SED model from Rowan-Robinson et al. (1997) together with the
quoted photometric redshifts to make corresponding $450\mu$m
maps. 

We find no correlation between the observed and simulated maps at
$450\mu$m. 
At $850\mu$m a weak correlation was found to be due entirely to NICMOS 
277.211, which has the second-brightest predicted flux and appears to
lie on a positive deflection in the $850\mu$m map. 
After excluding this galaxy, we could find no correlations in any 
subsets of the remaining sample. 
The galaxies with the brightest
$850\mu$m predictions are listed in Table \ref{tab:nicmos}. 


Interestingly, a KS test comparing the two
measurements of the brightest two objects with the total combined flux
distribution yields a probability of $0.074$ that the distributions
are the same. The mean flux is $1.0\pm 0.15$mJy, not far from the
predicted mean of $1.645$mJy. 
These both seem to suggest that there is real $850\mu$m flux at the
positions 
of the NICMOS galaxies with brightest $850\mu$m predictions. However  
for NICMOS 277.211 the $850\mu$m excess 
is due to a peak $\sim6''$ distant 
from the candidate source 
in both maps, but with a position consistent to around $\sim2''$ in
the two maps. So, the $850\mu$m flux may either 
be due to the NICMOS galaxy,
or may be due to this (presumably) unrelated apparent peak $6''$
away. The latter possibility may be ascribed 
to the effects of confusion at this flux density
level (Hogg 2000).  


\subsubsection{Summary of cross-correlations}

None of the catalogues considered so far yielded a
significant cross-correlation (barring individual point sources),
whether in the separate chop throw maps 
or in the combined maps, at either wavelength. Although the
probability of false negatives is not negligible, the fact that we
repeatedly fail to find a cross-correlation signal suggests the SCUBA
sources that comprise the sub-mJy fluctuations are on the whole
distinct from the ISO, VLA and Chandra populations. This is consistent
with the
lack of point sources in common between hard X-ray and sub-mm
sources, and with the source count model of Rowan-Robinson (2000), and
suggests that both the $850\mu$m background and point sources are 
dominated by star-forming galaxies at $z\stackrel{>}{_\sim}1$, and not 
by lower redshift galaxies or AGN. 

Elbaz et al. (2002) argue that a significant fraction of the
$140\mu$m extragalactic background light is due to the resolved
$15\mu$m galaxy population. Our cross-correlation results demonstrate 
that the $15\mu$m galaxies cannot also be responsible for
the $850\mu$m background. This confirms the expectation from source
count models (e.g. Rowan-Robinson 2001). 

The lack of statistical cross-correlations, and point-source
cross-identifications, 
is markedly at odds with the presence of the SCUBA sources
having near neighbours from 
multi-wavelength populations. For example, we have already discussed
the neighbouring $1.4$GHz populations; in figure
\ref{fig:scubahdf_sources} we plot the positions of the SCUBA sources
against the VLA blank fields (Richards et al. 1999) and the
Chandra-detected VROs (Alexander et al. 2001). Despite 
the apparent presence of clear SCUBA
cross-identifications with these Chandra and/or VLA sub-populations,
one SCUBA source (HDF850.1) is demonstrably identified with neither.  


%

\section{Conclusions}\label{sec:conclusions}
Our publicly-available SCUBA map of the Hubble Deep Field resolves
a substantial fraction 
%
%
%
%
%
of the extragalactic sub-mm background. 
At least half the
sources appear to be at $z\stackrel{>}{_\sim}1$, based on our
preliminary identifications. 
The lack of statistical 
cross-correlation signals with ISO, VLA or Chandra sources implies 
that the sources detected in these surveys are different populations
and/or at different redshifts, in turn implying the sub-mm 
galaxies at these flux densities are mainly high-redshift
($z>1$) galaxies with bolometric luminosities dominated by star
formation. We infer that the $\mu$Jy radio population lies
predominantly at $z<1$, in agreement with existing optical
spectroscopy, and that the populations dominating the hard X-ray
background contribute $<15\%$ of the sub-mm extragalactic background
light. 
Only two out of eight sub-mm sources are robustly identified with VLA
sources; radio pre-selection (e.g. Barger et
al. 2000, Chapman et 
al. 2001a,b) would therefore
have recovered $\geq2$ out of the $8$ SCUBA point 
sources in the HDF. Nevertheless, five of the eight sub-mm sources
have radio sources much closer than would be expected by chance, but
which are still not close enough to be physically 
identified with the sub-mm emission (provided we have not
underestimated the astrometric uncertainties). Millimetre-wave
interferometry has confirmed this is the case in the source HDF850.1, 
with a lensed near-infrared counterpart (Dunlop et al. 2002).  
This raises the interesting possibility 
that not all radio sources associatied with sub-mm galaxies are
responsible for the far-infrared emission. 
Unambiguous identifications have so far almost exclusively been
obtained using 
interferometric mm-wave follow-ups of brighter sub-mm sources,
reinforcing the strategy of 
wide-area, shallow sub-mm surveys for the study of the resolved sub-mm 
point source population
(e.g. Scott et al. 2002, Fox et al. 2002).

\section*{Acknowledgements}
This work was
supported by PPARC (grant 
number GR/K98728).
JD acknowledges the enhanced research time awarded by a PPARC Senior
Fellowship. 
The James Clerk Maxwell
Telescope is operated on behalf of the Particle Physics and Astronomy
Research Council of the United 
Kingdom, the Netherlands Organisation for Scientific Research and the
National Research Council of 
Canada. 
The authors acknowledge the data analysis facilities provided by the
Starlink Project which is run by CCLRC on behalf 
of PPARC. 

\section*{Appendix: Noise-weighted source extraction}

At every point $(i,j)$ in the image we wish to determine the best
fit (minimum $\chi^2$) point source. Suppose the point spread function
is $P(x,y)$, the image 
signal is $S(i,j)$, and the image noise is $N(i,j)$. The $\chi^2$ at
position $i,j$
is
\begin{equation}\label{eqn:a1}
\chi^2 (i,j) = \sum_{x,y}
                        \left( \frac{S(i-x,j-y)-F P(x,y) }{N(i-x,j-y)}\right)^2
\end{equation}                                                                
where $F$ is the best-fit flux at position $(i,j)$. If we minimise the
$\chi^2$ 
with respect to $F$ we obtain
\begin{equation}\label{eqn:a2}
\frac{d\chi^2}{dF} = -2\sum_{x,y} 
                  \frac{S(i-x,j-y)-F P(x,y)}{N(i-x,j-y)^2}
                  P(x,y)  .
\end{equation}
Setting this to zero and rearranging gives
\begin{equation}\label{eqn:a3}
F(i,j) = 
\frac{\sum_{x,y} S(i-x,j-y) W(i-x,j-y) P(x,y)}
     {\sum_{x,y} W(i-x,j-y) P(x,y)^2 }
\end{equation}
where $W=1/N^2$ can be thought of as the weights. 

This is equivalent to a convolution. 
We obtain the result that the best fit flux image is 

\begin{equation}\label{eqn:a4}
F = \frac{(SW)\otimes P}{W\otimes P^2}
\end{equation}
where $\otimes$ denotes a convolution.                 
Propagating the errors on F, we obtain
\begin{equation}\label{eqn:a5}
(\Delta F)^2 = \frac{1}{W\otimes P^2}
\end{equation}
Note that a signal - to - noise map here ($F/\Delta F$) is not the
same as a $\chi^2$ map: 
\begin{equation}\label{eqn:a6}
\frac{F}{\Delta F} = \frac{(SW)\otimes P}{\sqrt{W\otimes P^2}}
\end{equation}
whereas 
\begin{eqnarray}\label{eqn:a7}
\chi^2 & = & [(S/N)^2 \otimes (0\times P+1)]\nonumber \\ 
       &   & +   F^2 (W \otimes P^2) - 2 F [(SW) \otimes P]\\
       & = & [(S/N)^2 \otimes (0\times P+1)] 
-   F [(SW) \otimes P]\nonumber 
\end{eqnarray}
(the first step multiplies out eqn \ref{eqn:a1}, and the last step is
obtained by using the 
expression for F, eqn \ref{eqn:a4}).

This method of source extraction is now also being used on the $1''$
drizzling footprint images made for the ongoing $8$ mJy $850\mu$m
survey by ourselves (Scott et al. 2002, 
Fox et al. 2002).


\begin{thebibliography}{}
\bibitem{} Abergel, A., Andr\'{e}, P., Bacmann, A., et al., 1999, in
The Universe as seen by ISO, ESA-SP 427
\bibitem{} Alexander D.M., Vignali C., Bauer F.E., Brandt W.N., 
Hornschemeier A.E., Garmire G.P., Schneider D.P., 2001, AJ, 122, 2156
\bibitem{} Almaini, O., et al., 2002, MNRAS, submitted (astro-ph/0108400)
\bibitem{} Archibald, E.N., Wagg, J.W., Jenness, T., 2000, JAC
document SCD/SN/002, currently available at 
{\tiny\tt http://www.jach.hawaii.edu/JACdocs/JCMT/SCD/SN/002/index.html}
\bibitem{} Aussel, H., Cesarsky, C.J., Elbaz, D., Starck, J.L., 1999,
A\&A, 342, 313
\bibitem{} Barger A.J., Cowie L.L., Richards E.A., 2000, AJ, 119, 2092
\bibitem{} Barger, A., Cowie, L.L., Sanders, D.B., 1999,
ApJL, 518, 5
\bibitem{} Barger, A., Cowie, L.L., Smail, I., Ivison, R.J., 
Blain, A.W., Kneib, J.-P., 1999b, AJ, 117, 2656
\bibitem{} Barger, A.J., Cowie, L.L., Trentham, N., Fulton, E., Hu,
E.M., Songalia, A., Hall, D., 1999c, AJ, 117, 102
\bibitem{} Bertin E., Arnouts S., 1996, A\&AS, 117, 393
\bibitem{} Blain, A., 1999, in `Photometric Redshifts and High
Redshift Galaxies', R J Weymann, L J Storrie-Lombardi, M Sawicki, R J
Brunner eds. PASP conf series vol. 191, p. 255-264 (1999)
(astro-ph/9906141)
\bibitem{} Blain, A.W., Ivison, R.J., Kneib, J.-P., Smail, I., 1999,
in `The Hy-Redshift Universe: galaxy formation and evolution at high
redshift', eds. A. J. Bunker \& W. J. M. van Breughel, ASP 
conference vol. 193 (2000). ASP: San Francisco, p. 246-249
(astro-ph/9908024)
\bibitem{} Blain, A., Longair, M., 1993, MNRAS, 264, 509
\bibitem{} Borys, C., Chapman, S., Halpern, M., Scott, D., 2001, 
preprint (astro-ph/0107515)
\bibitem{} Brandt W.N., et al., 2001, AJ, 122, 1
\bibitem{} Carilli, C.L., Yun, M.S., 1999, ApJL, 513, 13
\bibitem{} Carilli, C.L., Yun, M.S., 2000, ApJL, 539, 1024
\bibitem{} Chapman S.C., Richards E.A., Lewis G.F., Wilson G., Barger A.J., 2001a, 2001, ApJL, 548, 147
\bibitem{} Chapman S.C., Lewis G.F., Scott D., Borys C., Richards E.A., 2001b, astro-ph/0111157 
\bibitem{} Cohen J.G., Cowie L.L., Hogg D.W., Songaila A., Blandford R.,
 Hu E.M., Shopbell P., 1996, ApJ, 471, L5
\bibitem{} Cohen J.G., Hogg D.W., Blandford R., Cowie L.L., Hu E., Songaila 
A., Shopbell P., Richberg K., 2000, ApJ, 538, 29
\bibitem{} Condon J.J., 1992, ARA\&A, 30, 575
\bibitem{} Condon, J.J.,  Kaplan, D.L., 1998, ApJS, 117, 361
\bibitem{} Conti, A., Kennefick, J.D,. Martini, P., Osmer, P.S., 1999, 
AJ, 117, 645
\bibitem{} Cooray A.R., 1999, New Astronomy, 4, 377
\bibitem{} D\'{e}sert, F. -X., Puget, J. -L., Clements, D.L.,
P\'{e}rault, M., Abergel, A., Bernard, J.-P., Cesarsky, C.J., 1999,
A\&A, 342, 363
\bibitem{} Downes A.J.B., Peacock J.A., Savage A., Carrie D.R., 1986, MNRAS,
218, 31
\bibitem{} Downes, D., et al., 1999, A\&A, 347, 809
\bibitem{} Dunlop, J., et al., 2002, preprint, astro-ph/0205480
\bibitem{} Eales, S., et al., 1999, ApJ, 515, 518
\bibitem{} Eales, S., Lilly, S., Webb, T., Dunne. L., 
Gear, W., Clements, D., Yun, M., 2000, AJ, 120, 2244
\bibitem{} Efstathiou, A., Rowan-Robinson, M., Siebenmorgen, 2000, 
MNRAS, 313, 734
\bibitem{} Elbaz, D., Cesarsky, C., Chanial, P., Aussel, H.,
Franceschini, A., Fadda, D., Chary, R,. 2002, A\&A, in press
\bibitem{} Fixen D.J., et al., 1998 ApJ, 508, 123
\bibitem{} Fox, M., et al., 2002, MNRAS, in press (astro-ph/0107585)
\bibitem{} Fruchter, A., Hook, R.N., 1997, in 
Proc. SPIE Vol. 3164, p. 120-125, Applications of
Digital Image Processing XX, Ed. Andrew G. Tescher. 
\bibitem{} Garrett, M.A., de Bruyn, A.G., Giroletti, M., Baan, W.A.,
Schilizzi, R.T., 2000, A\&A, 361, L41
\bibitem{} Gautier, T.N., Boulanger, F., P\'{e}rault, M., Puget, J.L., 
1992, AJ, 103, 1311
\bibitem{} Griffiths R.E., Ptak A., Feigelson E.D., Garmire G., Townsley L.,
 Brandt W.N., Sambruna R., Bregman J.N., 2000, Science, 290, 1325
\bibitem{} Guiderdoni, B., Bouchet, F.R., Puget, J.L., Lagache, G.,  
Hivon, H., 1998, Nature 390, 257 
\bibitem{} Herbstmeier, U., \'{A}br\'{a}ham, P., Lemke, D., Laureijs,
R.J., Klass, U., Mattila, K., Leinert, C., Surace, C., Kunkel, M.,
1998, A\&A, 332, 739
\bibitem{} Hogg, D.W., 2000, preprint astro-ph/0004054
\bibitem{} Holland, W.S., et al. 1999, MNRAS, 303, 659
\bibitem{} Hornschemeier, A.E., et al., 2000, ApJ, 541, 49
\bibitem{} Hughes, D., Serjeant, S., Dunlop, J., Rowan-Robinson, M.,
Blain, A., Mann, R.G., Peacock, J., Efstathiou, A., Gear, W., Oliver,
S., Lawrence, A., Longair, M., Goldschmidt, P., Jennes, T., 1998,
Nature, 394, 241
\bibitem{} Jarvis J.F., Tyson J.A., 1981, AJ, 86, 476
\bibitem{} Jarvis, R.M., MacAlpine, G.M., 1998, AJ, 116, 2624
\bibitem{} Jenness, T., 2000, JCMT Technical Report 84,
{\tiny\tt http://www.jach.hawaii.edu/JACdocs/JCMT/tr/001/84/tr0084.html}
\bibitem{} Lagache, G., Abergel, A.,  Boulanger, F., et al., 1999, 
A\&A 344, 322
\bibitem{} Lagache, G, Puget, J.L., 2000, A\&A, 355, 17
\bibitem{} Lanzetta K.M., Yahil A., Fernandez-Soto A., 1996, Nat, 381, 759
\bibitem{} Lilly, S.J., et al., 1999a, ApJ, 518, 641
\bibitem{} Lilly, S.J., et al., 1999b, in ``The formation of galactic
bulges'' eds. by C.M. Carollo, H.C. Ferguson, R.F.G. Wyse. Cambridge,
U.K. ; New York. 
Cambridge University Press, 1999. (Cambridge contemporary astrophysics), p.26
(astro-ph/9903157)
\bibitem{} Madau P., Ferguson H.C., Dickinson M.E., Giavalisco M., Steidel 
C.C., Fruchter  A., 1996, MNRAS, 283, 1388 
\bibitem{} Mann, R.G., et al., 1997, MNRAS, 289, 482,
\bibitem{} Mann, R.G., et al., 2002, MNRAS, in press
\bibitem{} Peacock et al., 2000, MNRAS, 318, 535
\bibitem{} Phillips A.C., Guzman R., Gallego J., Koo D.C., Lowenthal J.D.,
 Vogt N.P., Faber S.M., Illingworth G.D.,1997, ApJ, 489, 543
\bibitem{} Puget, J.L., et al., 1996, ApJL, 305, 5
\bibitem{} Richards E.A., 1999, ApJ, 513, L9 
\bibitem{} Richards, E., 2000, ApJ, 533, 611
\bibitem{} Richards, E.A., Kellermann, K.I., Fomalont, E.B., Windhorst, R.A., 
Partridge, R.B., 1998 AJ, 116, 1039
\bibitem{} Richards, E., et al., 1999, ApJL, 526, 73
\bibitem{} Rowan-Robinson, M., et al., 1997, MNRAS, 289, 490
\bibitem{} Rowan-Robinson, M., 2001, ApJ, 549, 745
\bibitem{} Rutledge R.E., Brunner R.J., Prince T.A., Lonsdale C.,
2000, ApJS, 131, 335
\bibitem{} Schlegel, D.J., Finkbeiner, D.P., Davis, M, 1998, ApJ, 500, 
525
\bibitem{} Scott, S., et al., 2002, MNRAS, in press (astro-ph/0107446)
\bibitem{} Serjeant, S., et al., 1997, MNRAS, 289, 457
\bibitem{} Serjeant, S., 2002 in preparation
\bibitem{} Serjeant, S., et al., 2002 in preparation
\bibitem{} Silva L., Granato G.L., Bressan A., Danese L., 1998, ApJ, 509, 103
\bibitem{} Simpson, C., Rawlings, S,. Lacy, M., 1999, MNRAS, 306, 828
\bibitem{} Smail, I., Ivison, R.J., Blain, A.W., 1997, ApJL, 490, 5
\bibitem{} Smail I., Ivison R.J., Blain A.W., Kneib J.-P., 1998, ApJ, 507, L21
\bibitem{} Smail, I,. Ivison, R.J., Kneib, J.-P., Cowie, L.L., 
Blain, A.W., Barger, A.J., Owen, F.N., Morrison, G., 1999, 
MNRAS, 308, 1061
\bibitem{} Sutherland W, Saunders W., 1992, MNRAS, 259, 413
\bibitem{} Thompson. R.I., 
Weymann, R.J., Storrie-Lombardi, L.J., 2000, ApJ, in press
(astro-ph/0008276) 
\bibitem{} Williams, R.E., et al., 1996, AJ, 112, 1335
\bibitem{} Wright, E.L., 1998, ApJ 496, 1

\end{thebibliography}
\end{document}